\documentclass[lettersize,journal]{IEEEtran}
\usepackage[dvipsnames, svgnames, x11names]{xcolor}
\usepackage{amsmath,amsfonts}
\usepackage{algorithmic}
\usepackage{algorithm}
\usepackage{array}
\usepackage[caption=false,font=footnotesize,labelfont=sf,textfont=sf]{subfig}
\usepackage{textcomp}
\usepackage{stfloats}
\usepackage{url}
\usepackage{verbatim}
\usepackage{graphicx}
\usepackage{cite}
\usepackage{multirow}
\usepackage[inline]{enumitem}
\usepackage{colortbl}

\definecolor{pink}{rgb}{0.99, 0.35, 0.34}
\definecolor{lightpink}{rgb}{0.99, 0.53, 0.52}
\definecolor{green}{rgb}{0.14, 0.33, 0.64}
\definecolor{lightgreen}{rgb}{0.73, 0.92, 0.99}
\definecolor{light}{rgb}{0.99, 0.69, 0.57}
\begin{document}

%\title{Silence and Speech Anti-spoofing: Impact, Reasons and Applications}
\title{The Impact of Silence on Speech Anti-Spoofing}

%\author{123,~\IEEEmembership{Student,~IEEE,}
%\author{Yuxiang Zhang,~\IEEEmembership{Student Member,~IEEE,} Zhuo Li,~\IEEEmembership{Student Member,~IEEE,} Jingze Lu,~\IEEEmembership{Student Member,~IEEE,}\\ Hua Hua,~\IEEEmembership{Student Member,~IEEE,}  Wenchao Wang~\IEEEmembership{Member,~IEEE,} Pengyuan Zhang~\IEEEmembership{Member,~IEEE,}
\author{Yuxiang Zhang, Zhuo Li, Jingze Lu, Hua Hua,  Wenchao Wang, Pengyuan Zhang
        % <-this % stops a space
%\thanks{This paper was produced by the IEEE Publication Technology Group. They are in Piscataway, NJ.}% <-this % stops a space
%\thanks{Manuscript received April 19, 2021; revised August 16, 2021.}}
\thanks{Manuscript received November 24, 2022; revised June 21, 2023; accepted August 03, 2023. Date of current version August 11, 2023. This work is partially supported by the National Key Research
	and Development Program of China (No. 2021YFC3320103). The associate editor coordinating the review of this manuscript and approving it for publication was Dr. Yu Tsao. (Corresponding author: Wenchao Wang and Pengyuan Zhang.)
	
	The authors are with Key Laboratory of Speech Acoustics and Content Understanding, Institute of Acoustics, Chinese Academy of Sciences, Beijing 100190, China and also with University of Chinese Academy of Sciences, Beijing 100049, China (e-mail: \{zhangyuxiang, lujingze, huahua, wangwenchao, zhangpengyuan\}@hccl.ioa.ac.cn and li\_zhuo@foxmail.com).}}

% The paper headers
\markboth{Journal of \LaTeX\ Class Files,~Vol.~14, No.~8, August~2021}%
{Yuxiang Zhang \MakeLowercase{\textit{et al.}}: The Impact of Silence in Anti-Spoofing}

% \IEEEpubid{0000--0000/00\$00.00~\copyright~2021 IEEE}
% Remember, if you use this you must call \IEEEpubidadjcol in the second
% column for its text to clear the IEEEpubid mark.

\maketitle

\begin{abstract}
 The current speech anti-spoofing countermeasures (CMs) show excellent performance on specific datasets. However, removing the silence of test speech through Voice Activity Detection (VAD) can severely degrade performance. In this paper, the impact of silence on speech anti-spoofing is analyzed. First, the reasons for the impact are explored, including the proportion of silence duration and the content of silence. The proportion of silence duration in spoof speech generated by text-to-speech (TTS) algorithms is lower than that in bonafide speech. And the content of silence generated by different waveform generators varies compared to bonafide speech. Then the impact of silence on model prediction is explored. Even after retraining, the spoof speech generated by neural network based end-to-end TTS algorithms suffers a significant rise in error rates when the silence is removed. %Furthermore, class activation mapping (CAM) visualization also shows that the model emphasizes the silence and the low frequency part of features. The differences in the content of silence between bonafide and spoof speech, as well as between speech generated by TTS algorithms and speech generated by voice conversion (VC) algorithms are demonstrated by masking silence or non-silence. 
 To demonstrate the reasons for the impact of silence on CMs, the attention distribution of a CM is visualized through class activation mapping (CAM). Furthermore, the implementation and analysis of the experiments masking silence or non-silence demonstrates the significance of the proportion of silence duration for detecting TTS and the importance of silence content for detecting voice conversion (VC). Based on the experimental results, improving the robustness of CMs against unknown spoofing attacks by masking silence is also proposed. Finally, the attacks on anti-spoofing CMs through concatenating silence, and the mitigation of VAD and silence attack through low-pass filtering are introduced.
\end{abstract}


\section{Introduction}
\IEEEPARstart{W}{ith} the development of deep learning, text-to-speech (TTS) \cite{wang2017tacotron, kong2020hifi} and voice conversion (VC) \cite{kaneko2018cyclegan} technologies have made significant improvements. Malicious use of spoof speech poses a huge threat to society. Attacking automatic speaker verification (ASV) systems with spoof speech generated by TTS and VC algorithms is defined as logical access (LA) attacks. To improve the performance of LA spoof speech detection countermeasures (CMs), several challenges have been successfully organized. The biennial ASVspoof Challenge \cite{wu2015asvspoof, Kinnunen2017, todisco2019asvspoof, yamagishi21_asvspoof} leads the way in speech anti-spoofing development and provides standard datasets containing a variety of TTS and VC algorithms. Unified evaluation metrics such as the equal error rate (EER) and the minimum normalized tandem detection cost function (min t-DCF) \cite{kinnunen2020tandem} are also proposed. Based on these challenges, a large number of CMs with great performance have emerged. 

Traditional anti-spoofing CMs can be divided into two parts: classifier and feature extraction. A popular category of classifiers is based on convolutional neural network (CNN), of which the residual neural network (ResNet) \cite{lai2019assert, chen21b_asvspoof, li2021replay} and the light convolutional neural network (LCNN) \cite{lavrentyeva2019stc, wang21fa_interspeech, tomilov21_asvspoof} are commonly used. The input features of CNN based classifiers are two-dimensional features including short-time Fourier transform (STFT) spectrogram, constant Q transform (CQT) spectrogram, Mel-scale transform (MSTFT), as well as cepstral features such as constant Q cepstral coefficient (CQCC)~\cite{todisco2017constant} and linear frequency cepstral coefficient (LFCC)~\cite{sahidullah15_interspeech}. Recently, neural network (NN) classifiers that use raw audio as input features have emerged, such as RawNet2 \cite{9414234, tak21_asvspoof} and graph attention network based AASIST \cite{9747766}. %The network architecture search is also utilized in spoof speech detection \cite{ge21c_interspeech, ge21_asvspoof}. 
The state-of-the-art (SOTA) systems can reach an EER of around 1\% \cite{9747766, tak21_asvspoof}.

Although showing excellent performance on specific datasets, current spoof speech detection algorithms still suffer from some shortcomings, especially poor robustness and lack of interpretability. The latest challenges focused on improving the robustness of spoof speech detection systems. Due to interference from coding and transmission artifacts introduced by telecommunication systems and compression codecs, data from different domains make ASVspoof 2021~\cite{yamagishi21_asvspoof} more challenging. The Audio Deep synthesis Detection Challenge (ADD) \cite{9746939} also considered realistic scenarios, including various real-world noises and background music effects, concatenating of fake segments with real speech, and new TTS and VC algorithms. Many methods for enhancing robustness, such as data augmentation \cite{das21_asvspoof} and adversarial training~\cite{chen21_asvspoof}~\cite{9746164} are applied to spoof speech detection. The latest self-supervised model effectively improves the robustness of CMs in complex scenarios. Systems using wav2vec 2.0 \cite{baevski2020wav2vec} gain a huge performance boost in ASVspoof 2021 Deepfake (DF) task \cite{wang2021investigating}, ADD low-quality fake audio detection \cite{9747768} and partially fake audio detection \cite{9747605} compared to systems with hand-crafted acoustic features.
 
In addition to the issues above, the impact of silence has overshadowed speech anti-spoofing. In the detection of replay attacks, non-speech has a significant impact in ASVspoof 2017 dataset \cite{8461467, chettri2020dataset} and ASVspoof 2019 PA dataset~\cite{chettri19_interspeech}. While there was less research on silence in LA scenarios at that time, we found that silence had a significant impact on LA spoof speech detection \cite{zhang21da_interspeech}. Removing silence segments by voice activity detection (VAD), especially at the beginning and end of the speech, results in approximately doubling the EER of CMs. Although the impact of silence on anti-spoofing systems has gradually attracted the attention of researchers and has become a hidden track of ASVspoof 2021 Challenge~\cite{liu2022asvspoof}, the reasons for this impact are still lacking in exploration. Recent work~\cite{muller21_asvspoof, 9599559} finds that due to the uneven distribution of silence duration in the training and evaluation partitions of ASVspoof 2019 LA dataset, the systems trained are considered to be more concerned with silence than speech. Here we improve the silence duration to the proportion of silence duration and show that the impact of silence differs for different spoofing algorithms. There are also some efforts to explain the anti-spoofing systems \cite{9747476}, but the impact of silence on LA anti-spoofing systems and its reasons is essential to be explored in-depth.

%If the judgment of the countermeasure is based mainly on silence, it is doubtful that current anti-spoofing algorithms are \emph{trustworthy} \cite{rusak2020increasing}. Deep learning models make decisions by fitting training data \cite{biship2007pattern}. Since spoof speech detection is a binary classification task. The model is vulnerable to overfitting and learning irrelevant cues or artifacts during training \cite{chettri2020dataset}. Once the background noise of the speech changes, or a simple post-processing of the speech by adding or cutting silence is performed, CMs will have a significant judgment error \cite{chen2022hccl}. This is an important reason for the poor robustness of the present CMs.

In this paper, the impact of silence on speech anti-spoofing, especially on the detection of different spoofing algorithms, is explored in detail. First, the difference in the silence between bonafide and spoof speech is intuitively analyzed. On the one hand, for the speech generated by most TTS algorithms, the proportion of silence duration is significantly smaller than that of bonafide speech when silence segments are not processed. %It can be found that the proportion of silence duration of the speech generated by TTS algorithms is significantly lower than that of bonafide speech. 
On the other hand, although the speech generated by VC algorithms and some TTS algorithms is similar to bonafide speech in the proportions of silence duration, there are differences in the content of silence generated by different waveform generators. %Thus the content of silence has an impact mainly on the detection of the speech generated by VC algorithms. 
%The experimental results of removing and masking silence demonstrate these findings. 
%Based on the experimental results about the content of silence, we argue that masking silence can exclude the content of silence and retain more robust information about the duration and location of silence. And improving the robustness of CMs in complex situations through masking the content of silence is proposed. %The EER can be reduced by about 10\% on the ASVspoof 2021 DF dataset after masking silence. 
Then the impact of silence on speech anti-spoofing is explored by comparing the detection capabilities of CMs %on various spoofing algorithms 
before and after removing silence. The reasons for the difference in silence are analyzed from how silence is embedded in the bonafide or spoof speech.% And the change in model predictions before and after removing silence is also shown.
To further demonstrate the reasons for the impact of silence, the CM based on STFT and Squeeze-and-Excitation ResNet (SENet)~\cite{hu2018squeeze} is visually analyzed with class activation mapping (CAM) \cite{zhou2016learning}. %By comparing the similarities and differences in the attention distribution of CM before and after VAD, %the reasons can be analyzed from another perspective. 
%it can be found that the silence at the beginning and end of speech is extremely important for detecting TTS-generated speech. While the silence at the interval of speech is the area where the model focuses on in VC-generated speech. %This is consistent with the findings regarding the proportion of silence duration and the content of silence.
%Additionly, based on the experimental results about the content of silence, we argue that masking silence can exclude the content of silence and retain more robust information about the duration and location of silence. 
In addition, removing silence changes both the proportion of silence duration and the content of silence. So experiments with masking silence or non-silence are performed to analyze the impact of silence content.
And improving the robustness of CMs in complex situations by masking the content of silence is proposed.

%A popular method for visual analysis of models is class activation mapping (CAM) . Grad-CAM \cite{selvaraju2017grad} is an improvement of CAM, where weights are derived from gradient information to obtain the distribution of model attention. Grad-CAM allows visualization of CMs based on spectrogram feature and CNN. The Grad-CAM are used in \cite{cheng2019replay, halpern20_odyssey} to explain countermeasures against LA and replay spoofing attacks . Further it is possible to visually analyze the regions where the model focuses on the features. 

%Therefore, we consider that the low-frequency part, together with the interval silence, characterizes the naturalness information such as speech rhythm and prosody, and is the main basis for the model to distinguish the spoofed speech. Moreover, a detailed analysis of different algorithm types is conducted and it is found that different spoof algorithms have different silence patterns and different sensitivities to silence.

With a clear understanding of how and why silence impacts speech anti-spoofing systems, in turn, concatenating silence segments at the beginning and end of spoof speech can attack anti-spoofing systems. By comparing the attack results of different silence segments, it can also be found that the content of silence has a significant impact on CMs. 

Last but not least, a method for mitigating VAD and silence attacks is also proposed. %In the visual analysis, only the low frequency part below 1 kHz of spectrogram receives attention. 
The low-frequency part can represent the pitch and fundamental frequency of speech. It can be assumed that the low-frequency part, together with interval silence, characterizes naturalistic information such as speech rhythm and prosody. %, and is a more robust feature for distinguishing spoof speech. 
And it is demonstrated that the EERs of anti-spoofing CMs can be reduced by low-pass filtering in the face of the test speech after VAD. Combining low-pass filtering speech and full-frequency silence can mitigate the performance degradation caused by silence attacks.

Overall, based on our previous work \cite{zhang21da_interspeech}, this paper is further extended and makes the following contributions.

\begin{itemize}
	\item The main reasons for the impact of silence on speech anti-spoofing are analyzed: On the one hand, speech generated by most TTS algorithms has a lower proportion of silence duration if the silence segments are not processed. On the other hand, the content of silence in spoof speech generated by different waveform generators is different from bonafide speech.
	
	\item The impact of silence and the two reasons are demonstrated through the experiments before and after VAD, as well as masking silence and non-silence. And a visualization analysis is conducted. Based on the impact of silence content, improving the robustness of CMs in complex scenarios by masking silence is proposed.
	
	\item Due to the difference in the proportion of silence duration, the silence at the beginning and end of speech is used to attack CMs. And the content of silence is demonstrated to be an important basis for detecting VC-generated spoof speech from another perspective.
	
	\item Mitigating the impact of VAD and silence attacks by low-pass filtering is proposed. %Based on the visualization analysis, the CM focuses on the low-frequency part of the non-silence part in spectrograms.
\end{itemize}

\section{Reasons for the impact of silence}\label{section2}
Several studies have demonstrated the impact of silence in both LA and PA spoof speech detection \cite{8461467, chettri2020dataset, chettri19_interspeech, zhang21da_interspeech, muller21_asvspoof, liu2022asvspoof}. However, the reasons for the impact remain unexplained. In this section, the reasons for the impact of silence on the anti-spoofing systems are explained visually.

The analysis of the impact of silence and its causes is developed mainly based on ASVspoof 2019 LA dataset. ASVspoof 2019 LA dataset \cite{wang2020asvspoof}, which was released in ASVspoof 2019 Challenge \cite{todisco2019asvspoof}, has contributed to the development of speech anti-spoofing. And there is still plenty of work done with this dataset today. Therefore, analyzing the impact of silence based on this dataset is convenient for comparison with other works. In addition, the dataset contains a rich set of spoofing algorithms, as well as nowadays commonly used spoofing algorithms, such as Tacotron2 \cite{shen2018natural} and WaveNet\cite{oord2016wavenet}. And the dataset has detailed algorithm descriptions. So it is possible to analyze the impact of silence and reasons on different algorithms. % Due to these reasons, the analysis of the impact of silence and its reasons is developed based on the ASVspoof 2019 LA dataset. 
The acoustic models and waveform generators in ASVspoof 2019 LA dataset are summarized in Table \ref{tab:data}.

\begin{table}[ht]
	\caption{The summary of ASVspoof 2019 LA dataset. \label{tab:data}}
	\centering
	\begin{tabular}{l l l}
		\hline
		Algorithms & Acoustic and conversion model & Waveform generator\\
		\hline
		A01 & AR RNN & WaveNet \\
		A02 & AR RNN & WORLD \\
		A03 & Feed-forward network & WORLD \\
		A04 & CART & Waveform concat. \\
		A05 & VAE & WORLD \\
		A06 & GMM-UBM & Spectral filtering \\
		\hline
		A07 & RNN & WORLD + GAN \\
		A08 & AR RNN & Neural source filter \\
		A09 & RNN & Vocaine (SPSS) \\
		A10 & Tacotron2 (AR RNN + CNN) & WaveRNN \\
		A11 & Tacotron2 (AR RNN + CNN) & Griffin-Lim (SPSS) \\
		A12 & RNN & WaveNet \\
		A13 & Feed-forward network & Spectral filtering \\
		A14 & RNN & STRAIGHT (SPSS) \\
		A15 & RNN & WaveNet \\
		A16 & CART & Waveform concat. \\
		A17 & VAE & Waveform filtering \\
		A18 & Linear & MFCC vocoder\\
		A19 & GMM-UBM & Spectral filtering \\
		\hline
	\end{tabular}
\end{table}

\subsection{The Proportion of Silence Duration}
The uneven distribution of the duration of silence in ASVspoof 2019 LA dataset was visualized in \cite{muller21_asvspoof}. However, due to the various lengths of speech, the simple statistics of the average length of silence cannot fully represent the distribution of silence. Therefore, the proportion of silence duration for each utterance is calculated. The proportion of silence is defined as the proportion of the silence frames compared to the total frames in an utterance and has been applied to speech emotion recognition \cite{atmaja2020effect}. The proportion of silence duration in an utterance can be calculated as:
$$
P=N_s/N_t
$$
where $N_s$ is the number of frames categorized as silence and $N_t$ is the total number of frames in the utterance. Each frame is categorized as silence or not by WebRTC VAD \cite{webrtcvad} with frame length of 10 ms.

\begin{figure}[htbp]
\centering
\includegraphics[width=1.0\linewidth]{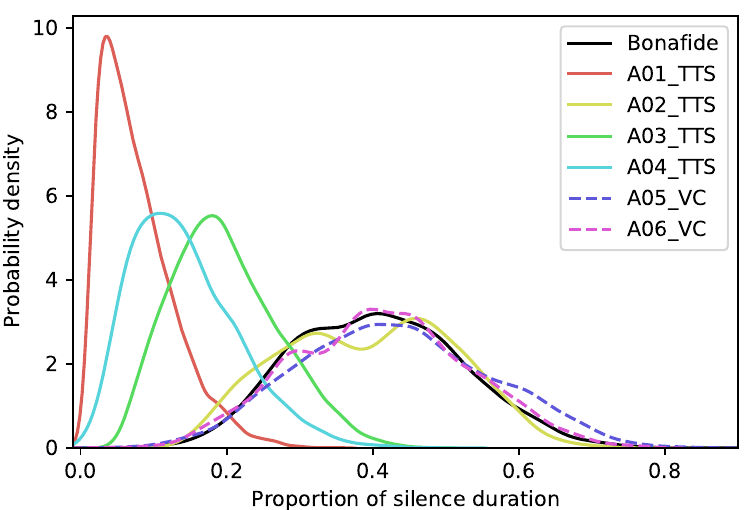}%
\caption{Probability density plot of the proportion of silence duration for ASVspoof 2019 LA training and development set.}
\label{fig_portion_td}
\end{figure}

\begin{figure}[htbp]
	\includegraphics[width=1.0\linewidth]{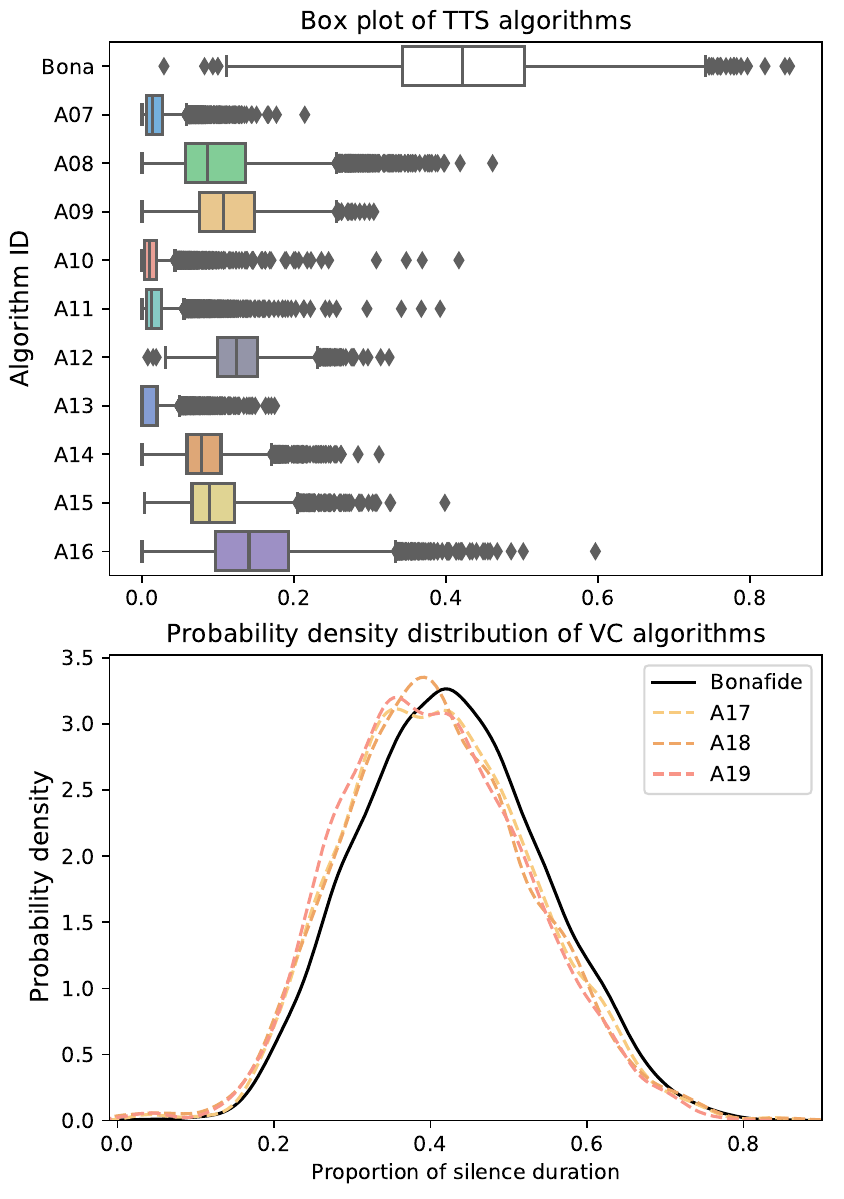}
	\caption{Box plot of the proportion of silence duration for TTS algorithms, as well as probability density plot for VC algorithms in ASVspoof 2019 LA evaluation dataset.}
	\label{fig_portion_eval}
\end{figure}

Figure \ref{fig_portion_td} shows the probability density distribution %boxplot
of the proportion of silence duration for different algorithms in ASVspoof 2019 LA training and development partitions. Since the algorithms of both subsets are the same, the proportion of silence duration for both subsets is counted together. %Except the A05 and A06 which are VC algorithms, the remaining four algorithms are TTS algorithms. 
The distribution of the proportion of silence duration between spoof speech generated by the TTS algorithms and the bonafide speech differs significantly except A02. As shown in Figure \ref{fig_portion_eval}, the difference in the distributions is more obvious in the evaluation dataset. %Since the proportion of silence duration of the speech generated by TTS algorithms is extremely small and the values are similar, it is difficult to obtain information clearly from the probability density plot. So it is displayed in a box plot instead. 
For bonafide speech, the proportion of silence duration is approximately Gaussian distributed, with a mean value of about $0.4$. 
The proportion of silence duration in the speech generated by TTS algorithms is significantly smaller and has a different distribution. 
In contrast, the distributions of the proportion of silence in the speech generated by VC algorithms are similar to that of bonafide speech in all subsets.

\begin{figure}[htbp]
	\centering
	\includegraphics[width=1.0\linewidth]{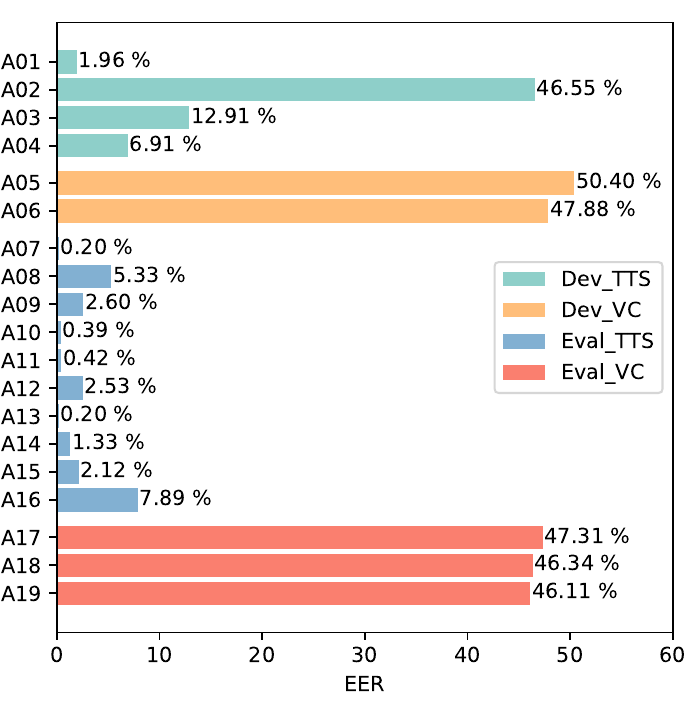}
	\caption{Results in terms of EER for ASVspoof 2019 LA Dev and Eval datasets based on the portion of silence.}
	\label{fig_portion_sil}
\end{figure}

To demonstrate this difference through data, the proportion of silence duration is directly used as the scores to calculate EER. Although a similar approach is used in \cite{liu2022asvspoof}, the analysis of different algorithms is missing. The results in Figure \ref{fig_portion_sil} can more significantly reflect the difference in the proportion of silence duration between the TTS-generated speech and the bonafide speech. Consistent with the distribution shown in the figures above, the proportion of silence duration has discrimination for A01, A03, and A04 in the development partition. And in the evaluation partition, the proportion of silence is effective in distinguishing all TTS algorithms. %The EER of all TTS-generated speech in the evaluation set is 3.02\%.
%However, the portion of silence is strongly correlated with the algorithm. For some TTS algorithms as well as all VC algorithms, the portion of silence is almost indistinguishable from bonafide speech.

\begin{figure}[htbp]
	\includegraphics[width=1.0\linewidth]{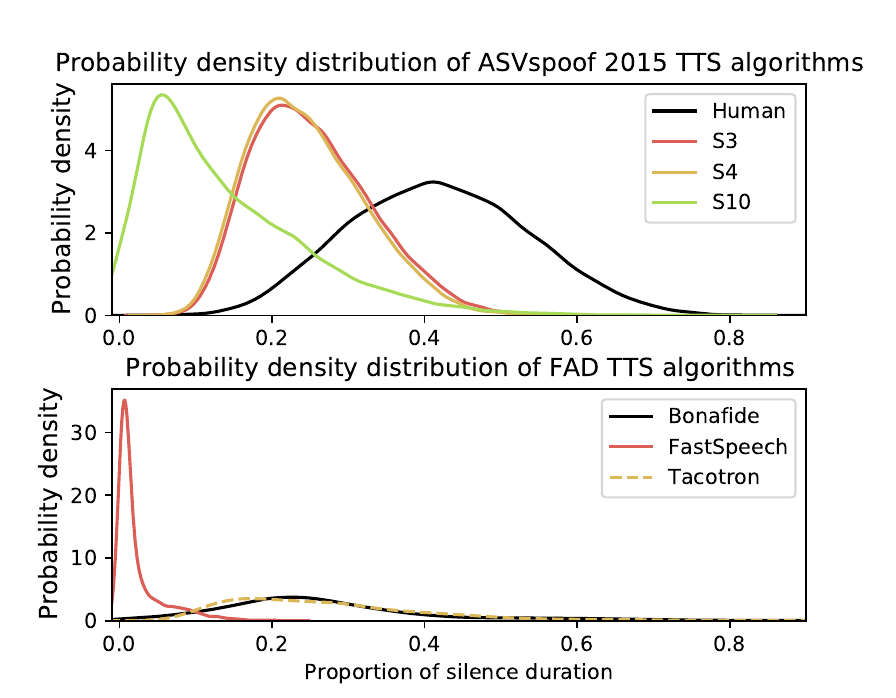}
	\caption{Probability density plot in ASVspoof 2015 and FAD datasets.}
	\label{fig_others}
\end{figure}

The proportion of silence duration for bonafide speech may vary depending on factors such as recording equipment and data processing. And the proportion of silence duration for spoof speech may also vary with training data and text analysis algorithms. This leads to the possibility that the patterns in ASVspoof 2019 LA dataset may fail in other datasets. To validate the generalizability of this hypothesis, the distribution of the proportions of silence duration in ASVspoof 2015~\cite{wu2015asvspoof} and fake audio detection dataset (FAD)~\cite{ma2023fad} are counted and shown in Figure \ref{fig_others}. In ASVspoof 2015, S3 and S4 are TTS algorithms based on hidden Markov model (HMM). S10 is a TTS algorithm implemented with the open-source MARY TTS system, which is similar to A04 and A16 in ASVspoof 2019 LA. The figure indicates that the proportion of silence duration of the speech generated by these three TTS algorithms is also lower compared to human speech. In FAD dataset, two TTS algorithms are implemented, based on Tacotron2 and FastSpeech. The proportion of silence duration of speech generated by FastSpeech is significantly smaller than bonafide speech. However, unlike A10 and A11 in ASVspoof 2019 LA, the proportion of silence duration of speech generated by Tacotron2 in FAD is similar to bonafide speech. It should be noted that the proportion of silence duration of bonafide speech in FAD dataset is lower than that in ASVspoof datasets. In general, the distribution of the proportion of silence duration of speech generated by TTS algorithms differs from that of bonafide speech. The proportion of silence duration of speech generated by TTS algorithms is smaller when the silence segments are not processed, and this is more prominent in ASVspoof 2019 LA. Exceptions may occur due to the process of data collection and algorithm implementation, such as A02 in ASVspoof 2019 and Tacotron in FAD.

The reason for the different proportion of silence duration can be inferred from the difference between the generation process of spoof speech and bonafide speech. The TTS systems usually consist of three parts: 1) Text analysis converts characters into phonemes or linguistic features; 2) Acoustic models generate acoustic features from linguistic features or characters/phonemes; 3) Vocoders generate waveform based on linguistic features or acoustic features \cite{tan2021survey}. All these parts lack the handling of silence, especially at the beginning and end. The acoustic models of TTS algorithms generate acoustic features based on text sequences and do not add silence between sentences. And in sentences, pauses are an important part of prosody \cite{10.5555/214780.214786}, and are strongly related to the fillers, speaker, style, etc \cite{betz:hal-02360611}. However, due to the rich variety of speech and text, it is difficult for current text analysis algorithms to model diverse and accurate pauses. In contrast, the bonafide speech is obtained by recording. And inevitably, the silence at the beginning and end is retained during the recording process. There are also inevitable pauses during inspiration and respiration \cite{10.5555/214780.214786}. Thus bonafide speech contains a higher proportion of silence duration than speech generated by TTS or TTS\_VC algorithms when the silence is not processed. Since the spoof speech generated by VC algorithms is based on bonafide speech, it also preserves a high proportion of silence duration.

\subsection{The Content of Silence}
For most TTS algorithms, the proportion of silence duration is significantly smaller. In contrast, for the spoof speech generated by VC algorithms, and some of TTS algorithms such as A02, the proportion of silence duration is similar to that of the bonafide speech. But silence still has an impact on the detection of these algorithms. So the proportion of silence duration may not be the only reason why silence has an impact on speech anti-spoofing. Also analyzed from the difference between the bonafide and spoof speech generation process, there are also differences in the content of silence. The content of the silence may include inspiration, swallowing, any laryngo-phonatory reflex or silent expiration~\cite{10.5555/214780.214786}, or breathing during the intervals of continuous speech~\cite{gao2022audio}. The silence from the inevitable breathing during natural pronunciation is so informative that can even be used in speaker identification \cite{zhao2017speaker}. Acoustic models in TTS usually do not generate acoustic features or generate features that are strongly correlated with the training data at pauses. Due to the inevitable loss in the waveform generator, the generated waveform at the silence is also defective. These reasons lead to differences in the silence content of spoof speech and the highly random silence content of bonafide speech. %Thus there are differences in the content of silence between bonafide and spoof speech. 
For spoofing algorithms in ASVspoof 2019 LA dataset, the content of silence can be broadly classified into three types according to the waveform generator: WORLD \cite{morise2016world} class, such as A02 and A05; other filtering based class, such as A06, A17, A19; and waveform concatenation class including A04 and A16.

\begin{figure}[htbp]
	\centering
	\includegraphics[width=1.0\linewidth]{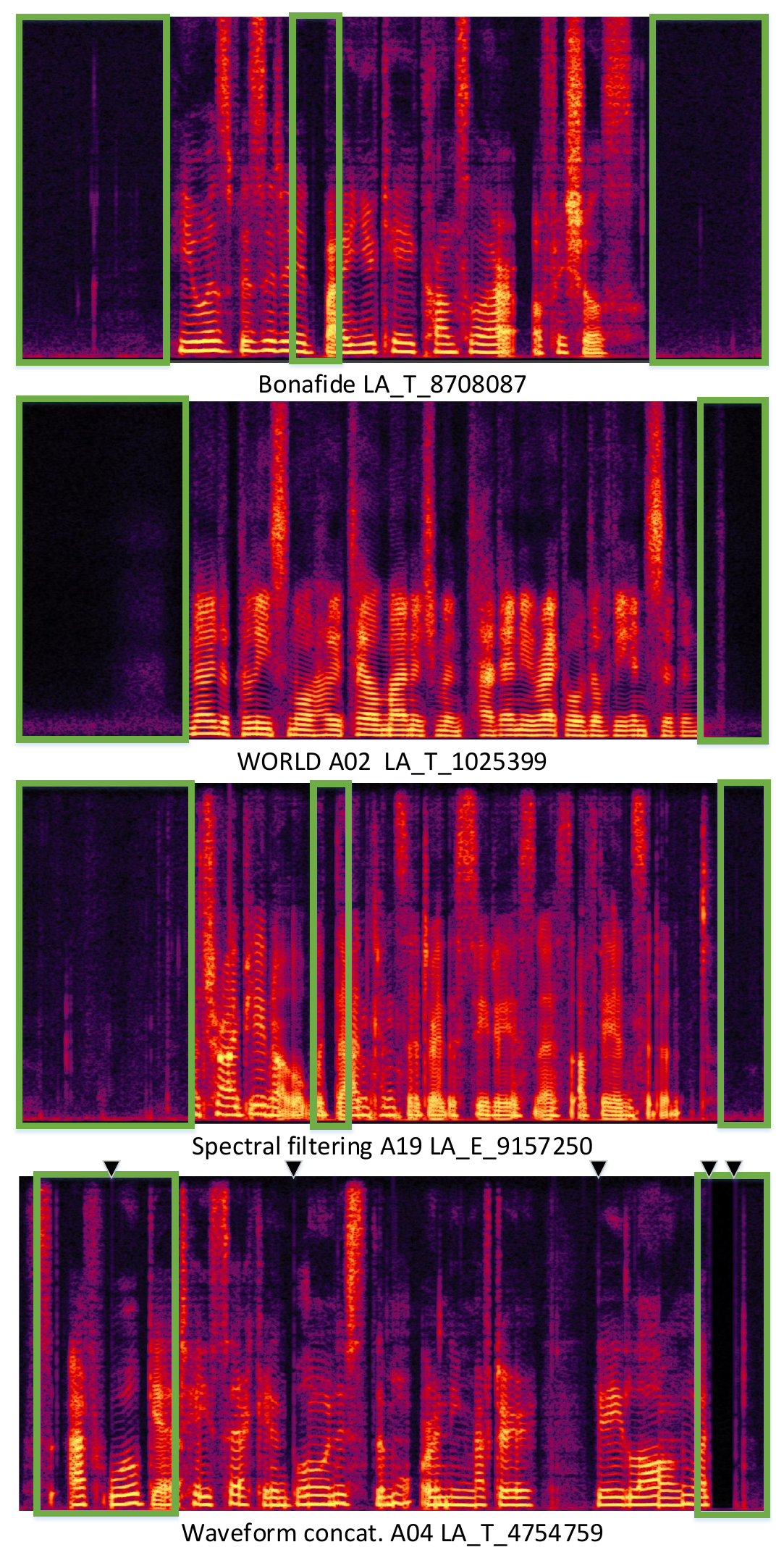}
	\caption{Spectrogram differences in the content of silence due to different waveform generators.}
	\label{fig_sil_content}
\end{figure}

As shown in Figure \ref{fig_sil_content}, the content of silence for all three classes of spoofing algorithms differs from that of the bonafide speech. For speech generated by TTS based on WORLD, its silence content lacks white noise and the speech interval compared with bonafide speech. Even though white noise and breathing sounds are present in VC algorithms based on WORLD, there are no sounds associated with the vocalization process, such as lip contact. For the VC algorithms based on spectral filtering, the silence content is most similar to that of bonafide speech. Especially A06 and A19 only convert the voice frames detected by VAD and do not change the non-speech frames \cite{wang2020asvspoof}. But there are still some differences, such as some vertical bands in the spectrogram that seem to be impulse noise, and the too-short interval. As for the spoof speech generated by waveform concatenation, silence is missing at the beginning and end of the speech. Due to the signal discontinuity caused by the concatenating, there is spectral broadening across all frequencies at the possible concatenation points. Four possible concatenation points are marked with triangles in Figure \ref{fig_sil_content}. 

\begin{figure}[htbp]
	\centering
	\includegraphics[width=1.0\linewidth]{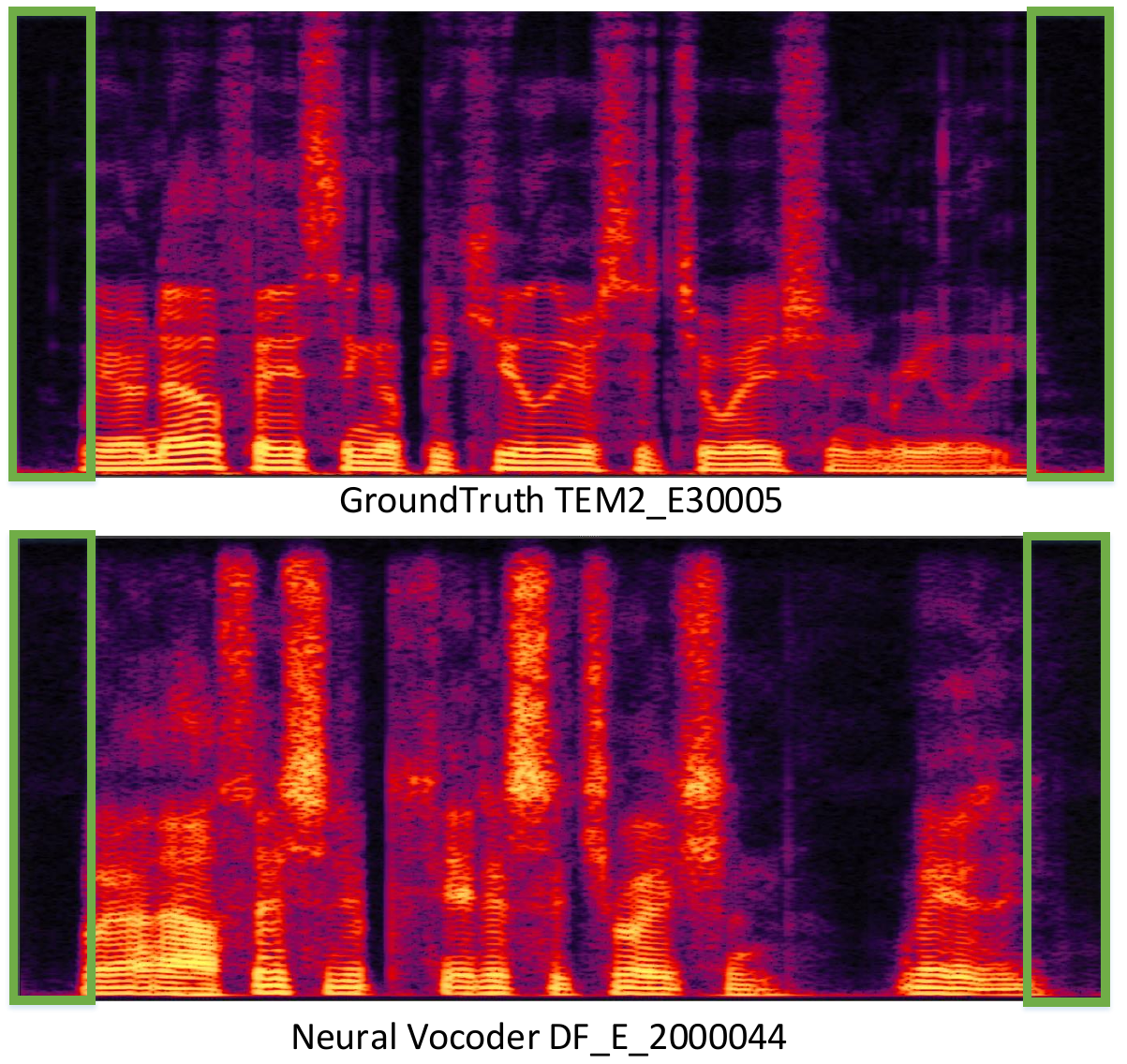}
	\caption{Difference in the content of silence of a speech in VCC 2020 and ASVspoof 2021 DF.}
	\label{fig_sil_content_vcc}
\end{figure}

To demonstrate the generalization of the content of silence, the comparison of a speech in Voice Conversion Challenge (VCC) 2020 \cite{Yi2020} and ASVspoof 2021 DF \cite{liu2022asvspoof} dataset is supplemented in Figure \ref{fig_sil_content_vcc}. The spoof speech is generated by a non-autoregressive neural vocoder. Even with the speech of the same content, the VC algorithm still generates speech with differences in the content of silence.

Collectively, the proportion of silence duration and the content of silence are the main differences between the bonafide and spoof speech. Silence becomes a \emph{shortcut} \cite{geirhos2020shortcut} to NN-based CMs since it is easier to discriminate between bonafide and spoof speech in the training set through these differences in silence than by differences in non-silence. They are also the reasons why silence has a great impact on anti-spoofing CMs, as shown below. In order to learn the difference between bonafide and spoof speech in non-silence, more training data may be required \cite{tyagi2013application}.

\section{Spoofing countermeasures}\label{section3}
To fully analyze the decision basis across different CMs, this section describes three CMs implemented in this paper. Detailed descriptions of the evaluation metrics and datasets are also provided.

\subsection{Features}
Two acoustic features are used as input for CNN-based CMs, including low-frequency STFT spectrograms for SENet and LFCC fed into LCNN. 

The STFT spectrogram is similar to the excellent system in ASVspoof 2019 \cite{lavrentyeva2019stc}, and is extracted with window length of 108 ms, hop length of 8.125 ms, and Fast Fourier Transform (FFT) points of 1,728. And the number of frames of the spectrogram features is fixed at 600. The shorter spectrograms are padded reflectively and the longer spectrograms are trimmed. The low-frequency features are demonstrated to have better robustness than the full spectrograms \cite{zhang21da_interspeech}. Features that pass through a finite impulse response (FIR) filter for data augmentation \cite{tomilov21_asvspoof, 9747768} have similar ideas to our low frequency features. Therefore, only the spectrograms of the low frequency part below 4 kHz are used as input features.

As one of the baseline systems for ASVspoof 2021, the 60-dimensional LFCC feature is similar to \cite{wang21fa_interspeech} and is extracted with window length of 20 ms, hop length of 10 ms, FFT of 1024 points, a linearly spaced triangle filter bank of 20 channels and delta plus delta-delta coefficients. In addition, the maximum frequency of extracted FFT spectrograms is also 4 kHz. To be consistent with the SENet system, the duration of speech is fixed to 6 seconds by reflection padding or trimming before extracting features.

As in \cite{9747766}, the end-to-end CM AASIST directly feeds raw speech into SincNet \cite{ravanelli2018speaker} to extract features. The length of the input speech is 4 s. If the speech is shorter it will be padded by repetition. If the speech is longer, a segment of speech is selected from a random starting point.

\subsection{Classifiers}
Two CNN-based classifiers and one end-to-end classifier are implemented to explore the impact of silence.

The SENet is a combination of ResNet with the Squeeze and Excite (SE) block \cite{hu2018squeeze}, which is one of the commonly used models for spoof speech detection. %The jump connection of ResNet deepens the depth of the network while preventing the gradient vanishing. The SE block adaptively captures the importance of each feature channel and explicitly models the interdependencies between channels by assigning weights. Features that are useful are boosted according to importance, while useless features are suppressed. Due to these advantages, 
The SENet can assign weights to the features of different channels through attention, thereby enhancing features that are more important for anti-spoofing. The SENet implemented here is SENet34. The angular margin based softmax loss (A-softmax) \cite{liu2017sphereface} is used as the loss function. Adam \cite{kingma2015adam} is adopted as the optimizer with $\beta_1=0.9$, $\beta_2=0.98$, $\epsilon=10^{-9}$ and weight decay $10^{-4}$. 

The LCNN is the same as the baseline for ASVspoof 2021~\cite{todisco2019asvspoof, wang21fa_interspeech}, with two Bi-LSTM layers added to the LCNN9 for pooling. A skip connection is added to the two Bi-LSTM layers. The size of the Bi-LSTM layers and the size of the LCNN output feature are both 128. The final fully connected layer projects the 128-dimensional embeddings to 2-dimensional outputs. The loss function and optimizer are the same as the SENet above.

The end-to-end anti-spoofing CM AASIST is the same as \cite{9747766}, where the encoder is based on six residual blocks. The graph attention layer and the graph pooling layer are similar to \cite{tak21_asvspoof}.  The graph attention layer uses heterogeneous attention mechanism and stack nodes to model artifacts across heterogeneous temporal and spectral domains. The max graph operation can improve the robustness of the end-to-end anti-spoofing system. The loss function is cross-entropy loss with weights of (0.1, 0.9) for spoof and bonafide classes. Adam optimizer with a learning rate of $10^{-4}$ and cosine annealing learning rate decay are utilized.

\subsection{Dataset and Metrics}
The training of models is based on the training set of ASVspoof 2019 LA database, whose spoof speech is generated by 6 different spoofing algorithms. The development partition contains the same algorithms as the training partition and is used for model selection and determining if the model is overfitting. The model with the lowest loss on the development set is selected as the model for evaluation. In evaluating of ASVspoof 2021 datasets, the models are trained with online data augmentation. The MUSAN\cite{musan2015} and RIRS\_NOISES\cite{ko2017study} datasets are used for data augmentation.

The evaluation data includes the evaluation partition of ASVspoof 2019 LA, ASVspoof 2021 LA, as well as ASVspoof 2021 DF datasets. The evaluation partition of ASVspoof 2019 LA includes 2 known algorithms (A16 and A19) and 11 unseen algorithms which are different from those in the training and development partitions. ASVspoof 2021 LA evaluation data includes speech over a variety of telephony systems, including Voice over IP (VoIP) and Public Switched Telephone Network (PSTN). The transfer of data across channels introduces nuisance variability that can arise in real-world application scenarios. The spoofing algorithms in the dataset are the same as ASVspoof 2019 LA evaluation partition. ASVspoof 2021 DF evaluation data is a collection of bonafide and spoof speech processed with different codecs typically used for media storage. The codec process creates distortion, which significantly lowers the performance of CMs. In addition, ASVspoof 2021 DF evaluation dataset contains spoofing attacks produced by more than 100 different spoofing algorithms. These factors make ASVspoof 2021 DF dataset close to the real scene, which is beneficial to evaluate the robustness of CMs.

The metric used here is EER, which is an evaluation metric used by both ASVspoof Challenge and ADD Challenge. EERs of different algorithms are also calculated and reported.

\section{Exploring the impact of silence}\label{section4}
This section explores the impact of silence on the CMs. The performance differences of the three CMs described in Section \ref{section3} before and after silence removal are analyzed. Through experiments, it is demonstrated that even after retraining, current CMs cannot work effectively without silence. The reasons for the impact of silence on the CMs are also explored visually through CAM. The significance of the proportion of silence duration for detecting TTS and the importance of silence content for detecting VC is demonstrated by masking silence and non-silence. Based on this experiment, we propose masking silence to improve the robustness of CMs in the face of out-of-distribution data such as ASVspoof 2021 DF. %Based on rich experimental results as well as detailed analysis, the proportion of silence duration and the content of silence proved to be important reasons for the impact of silence on speech anti-spoofing.

\subsection{Impact of Silence on Model Prediction}\label{subsec4.1}
It has been shown that silence, especially at the beginning and end of the speech, has a significant impact on CMs~\cite{zhang21da_interspeech, liu2022asvspoof,  muller21_asvspoof, 9599559}. Even after retraining using data without silence, CMs still have difficulty detecting spoof speech without silence. %Furthermore, we find that silence has different impact on the detection performance of different spoofing algorithms. 
In this section, the impact of silence on different spoofing attacks is analyzed by comparing the experimental results with or without silence during training and prediction. This leads to a preliminary demonstration of the reasons why silence has a significant impact on CMs.

\begin{table*}[htbp]
	\caption{Results in terms of EER/\% for algorithms in ASVspoof 2019 LA development (A01-A06) and evaluation (A07-A19) subsets under different training and prediction configurations. The EER of each spoof method is calculated separately.}\label{tab:VAD}
	\centering
	\setlength{\tabcolsep}{3.0pt}
	\begin{tabular}{ l | c | c c c c c c | c | c c c c c c c c c c c c c | r}
		\hline
		\textbf{Model} & Conf & A01 & A02 & A03 & A04 & A05 & A06 & Dev & A07 & A08 & A09 & A10 & A11 & A12 & A13 & A14 & A15 & A16 & A17 & A18 & A19 & Eval\\
		\hline
		\multirow{3}{*}{SENet} & \romannumeral1 & 0.24 & 0.51 & 0.24 & 0.48 & 0.54 & 1.02 & 0.55 & 0.19 & 1.77 & 0.06 & 0.37 & 0.53 & 0.08 & 0.16 & 0.08 & 0.27 & 0.35 & 2.50 & 1.24 & 1.48 & 1.14\\
		& \romannumeral2 & 21.3 & 3.30 & 7.97 & 29.2 & 5.49 & 3.06 & 14.8 & 31.8 & 18.1 & 9.64 & 44.1 & 54.9 & 31.2 & 21.9 & 13.5 & 28.1 & 23.7 & 7.55 & 9.91 & 5.07 & 25.5 \\
		& \romannumeral3 & 3.22 & \textbf{1.10} & \textbf{1.53} & \textbf{6.99} & 1.34 & 3.06 & 3.26 & 27.8 & 20.4 & \textbf{3.06} & \textbf{35.0} & \textbf{42.5} & 16.9 & 3.93 & 11.1 & 29.6 & 6.96 & 2.97 & 2.38 & 4.46 & 20.1 \\
		\hline
		\multirow{3}{*}{LCNN} & \romannumeral1 & 0.40 & 1.25 & 0.08 & 1.13 & 0.35 & 1.41 & 0.94 & 0.51 & 3.15 & 0.08 & 0.55 & 0.53 & 0.47 & 0.47 & 0.42 & 0.38 & 0.71 & 21.9 & 17.6 & 6.63 & 6.76 \\
		& \romannumeral2 & 15.4 & 16.6 & 9.93 & 38.8 & 17.4 & 1.85 & 18.5 & 47.8 & 9.89 & 3.03 & 45.3 & 37.3 & 32.3 & 37.0 & 28.1 & 24.9 & 34.2 & 16.6 & 22.8 & 5.13 & 28.7 \\
		& \romannumeral3 & 9.37 & \textbf{4.28} & \textbf{3.65} & \textbf{32.0} & 4.82 & 12.1 & 13.3 & 47.4 & \textbf{3.75} & \textbf{0.53} & \textbf{46.0} & \textbf{47.0} & 27.7 & 25.7 & 23.3 & 9.99 & 30.3 & 7.83 & 21.8 & 19.6 & 26.4 \\
		\hline
		\multirow{3}{*}{AASIST} & \romannumeral1 & 0.24 & 0.16 & 0.11 & 0.62 & 0.40 & 0.94 & 0.47 & 0.91 & 0.38 & 0.04 & 1.24 & 0.45 & 0.87 & 0.14 & 0.20 & 0.67 & 0.95 & 1.69 & 4.56 & 0.67 & 1.59 \\
		& \romannumeral2 & 4.04 & 0.03 & 0.11 & 12.2 & 1.34 & 4.20 & 5.45 & 45.2 & 1.34 & 0.67 & 51.3 & 23.2 & 45.9 & 13.9 & 9.14 & 37.5 & 12.4 & 5.49 & 9.18 & 2.71 & 23.8 \\
		& \romannumeral3 & 0.24 & \textbf{0.11} & \textbf{0.11} & \textbf{3.65} & 0.48 & 1.13 & 1.34 & 38.0 & \textbf{0.34} & \textbf{0.10} & \textbf{45.6} & \textbf{33.8} & 35.8 & 6.58 & 7.69 & 22.3 & 4.78 & 1.87 & 4.78 & 1.14 & 21.0 \\
		\hline
	\end{tabular}
	
\end{table*}

The three training and prediction configurations in the experiment are as follows, with the same VAD implementation as in Section \ref{section2}:

\begin{enumerate}[label={\roman*.}]
	\item The training and prediction of models are done using unprocessed speech.
	\item The models are trained using unprocessed speech, and prediction is done using speech without silence.
	\item The models training and prediction are done using the speech cut out of the silence by VAD.
\end{enumerate}

As shown in Table \ref{tab:VAD}, The performance of all models degrades dramatically when using speech without silence for prediction, regardless of whether the training was done with or without silence. By retraining models with speech without silence, the EERs of the predictions on the development set are significantly reduced. However, the models still struggle to effectively detect unknown spoofing attacks without silence in the evaluation partition.

Compared to the TTS algorithms (A01-A04, A07-A16), the VC algorithms (A05, A06, A17-A19) have less performance degradation under configuration \romannumeral2 \  and \romannumeral3. An exception is the difficulty of detecting A06 and A19 algorithms in the LFCC-LCNN system under configuration \romannumeral3. A possible explanation is that the LFCC feature is primarily concerned with spectral envelope information and may be more sensitive to speech silence connections. So removing the silence and then concatenating the speech further breaks the speech-silence connection within the speech, causing the LFCC features to fail to detect the speech generated by A06 and A19.

The VC algorithms are more difficult to detect under the condition that silence exists, while the TTS algorithms are more difficult to detect when silence is removed. This is consistent with the different proportions of silence duration shown in Section \ref{section2}. Among the TTS algorithms, A02, A03, A08, and A09 still have small EERs after removing silence. The common points between them \cite{wang2020asvspoof} are that the acoustic models are based on an autoregressive structure, while the vocoders are source filter-based models, such as WORLD~\cite{morise2016world}. The source filter model introduces white noise as an aperiodic excitation signal for speech waveform generation~\cite{8682298, 7178768}. A probable explanation is that these aperiodic excitation signals introduce silence information and the proportion of silence duration is higher among the TTS algorithm, making the CMs more focused on the non-silence part. The algorithm that is more impacted by silence in the development partition is the A04 based on waveform concatenating. A10 and A11 are algorithms with large performance degradation in the evaluated partitions. Both of them are based on Tacotron2 \cite{shen2018natural}, which converts character sequences into Mel-spectrograms frame by frame through recurrent neural network (RNN). This end-to-end acoustic model is an advanced acoustic model \cite{yang2022torchaudio} that generates high quality acoustic features and has a wide range of applications~\cite{cooper2020zero}. However, the features generated by this model lack silence at the beginning and end. As a result, CMs trained with unprocessed speech can detect such spoof speech by the duration of silence, whereas models trained on speech without silence are difficult to detect for these high quality unknown algorithms.

\begin{figure}[htbp]
	\centering
	\includegraphics[width=1.0\linewidth]{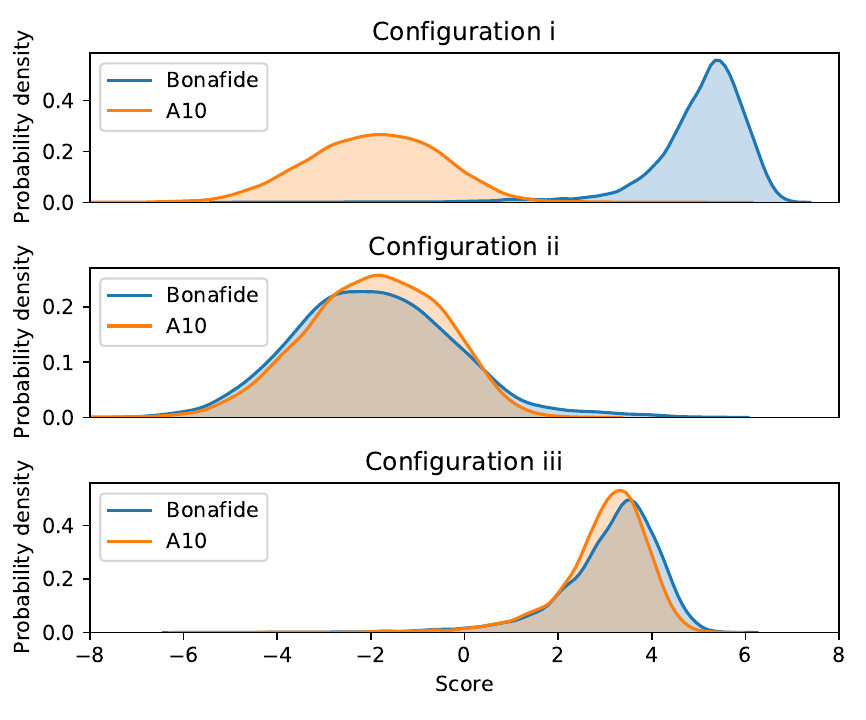}
	\caption{AASIST scores distribution of the speech generated by the A10 algorithm and the bonafide speech under the three configurations.}
	\label{fig_score}
\end{figure}

Because A10 has the most severe performance degradation, the distribution of the scores of the A10-generated speech and the bonafide speech are shown in Figure \ref{fig_score}. The scores are obtained through the AASIST system. When the threshold is set to zero, the false rejection rate for a bonafide speech is higher in configuration \romannumeral2. However, the false acceptance ratio of the spoof speech generated by A10 is significantly higher for the model trained with speech without silence. This indicates that the model trained with speech without VAD tends to classify the speech lacking silence as spoof speech. While the model trained with speech without silence has difficulty in classifying the spoof speech generated by unknown TTS algorithms.

\begin{table}[htbp]
	\caption{The EER/\% for ASVspoof 2021 DF (DF) and ASVspoof 2021 LA (LA) progress and evaluation datasets under different training and prediction configurations. \label{tab:vad_21}}
	\centering
	\begin{tabular}{c c c c c c}
		\hline
		\multirow{2}*{Model} & \multirow{2}*{Configurations} & \multicolumn{2}{c}{DF} & \multicolumn{2}{c}{LA}\\
		\cline{3-6}
		& & Prog & Eval & Prog & Eval \\
		\hline
		\multirow{3}{*}{SENet} & \romannumeral1 & 7.58 & 22.95 & 9.01 & 8.71 \\
		& \romannumeral2 & 36.58 & \textbf{39.84} & 39.14 & \textbf{42.42} \\
		& \romannumeral3 & 28.35 & 30.89 & 26.49 & 28.31 \\
		\hline
		\multirow{3}{*}{LCNN} & \romannumeral1 & 6.71 & 22.46 & 6.62 & 6.89 \\
		& \romannumeral2 & 34.03 & 35.61 & 34.55 & 35.75 \\
		& \romannumeral3 & 30.24 & 33.23 & 31.75 & 32.96 \\
		\hline
		\multirow{3}{*}{AASIST} & \romannumeral1 & 3.71 & 16.92 & 8.65 & 7.52 \\
		& \romannumeral2 & 28.66 & 31.20 & 31.09 & 28.66 \\
		& \romannumeral3 & 25.64 & \textbf{27.88} & 26.13 & \textbf{27.66} \\
		\hline
	\end{tabular}
\end{table}

The impact of different silence configurations on ASVspoof 2021 datasets is similar to that on ASVspoof 2019 datasets as shown in Table \ref{tab:vad_21}. Comparing the impact of different silence configurations, the worst performance is found for configuration \romannumeral2, in which training and testing do not match. The SENet model is the most affected by silence, while AASIST is the most robust. The performance of the three models is improved by retraining models with speech without silence. But the improvements are not obvious, and the LCNN model shows the least improvement.

\subsection{Exploring Silence through CAM}\label{sub4.2}
In order to visually analyze the reasons for the significant impact of silence on CMs, we visualized the CM based on the STFT spectrogram and SENet. The visualization analysis is performed by Grad-CAM \cite{selvaraju2017grad}, which derives the attention distribution of the trained model on the feature map from the gradient information. 

According to the analysis in subsection \ref{subsec4.1}, three spoofing algorithms of A09, A10, and A18 are used as examples for comparative analysis through CAM. The reasons why different TTS algorithms have different sensitivity to silence are analyzed through A09 and A10. The reasons why silence has a smaller impact on the VC algorithms are explained using A18. One speech from each of the three algorithms and bonafide speech is randomly selected. And the SENet model trained with configuration \romannumeral1 \enspace and \romannumeral3 \enspace is used for prediction. Grad-CAMs show the difference in the region where the model focuses on before and after removing silence.

\begin{figure}[htbp]
	\centering
	\includegraphics[width=1.0\linewidth]{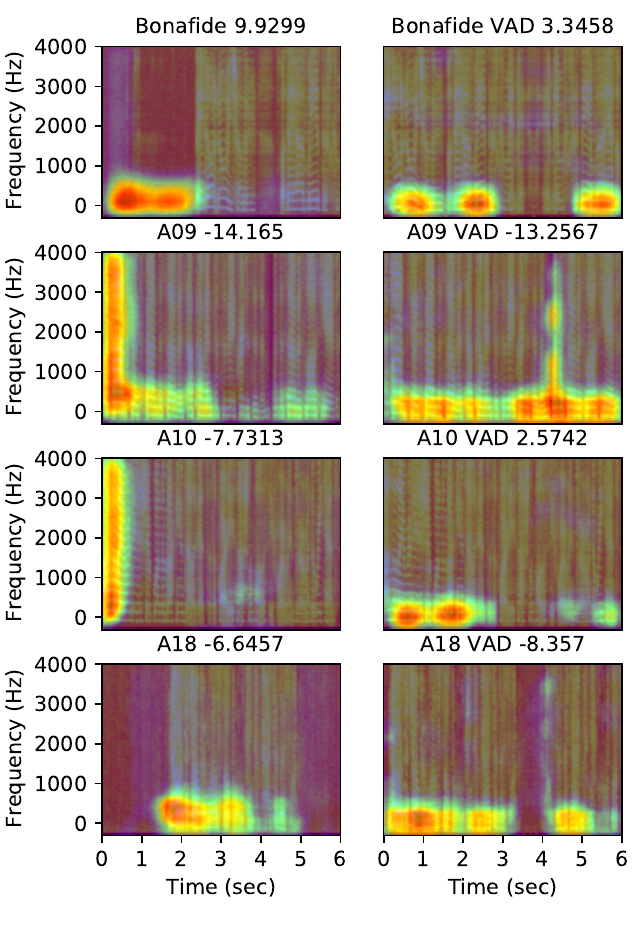}
	\caption{CAM heatmaps for bonafide speech and spoof speech generated by different algorithms. Left represents the attention distribution in configuration i, and the right represents the attention distribution in configuration iii. The names of algorithms are followed by the scores of the model prediction.}
	\label{fig_cam}
\end{figure}

As shown in Figure.\ref{fig_cam}, from top to bottom, heatmaps correspond bonafide speech LA\_T\_1563149, and spoof speech LA\_E\_1023808, LA\_E\_1002474, LA\_E\_1323026 generated by A09, A10, and A18. As shown on the left, the model trained with unprocessed speech is mainly concerned with the presence of silence at the beginning of speech when detecting the speech generated by TTS algorithms such as A09 and A10. Since silence exists at the beginning of both, the distinction between the bonafide speech and the spoof speech generated by VC algorithms is mainly based on the spectrogram of frequency below 1 kHz in the non-silence part. For the model trained with speech without silence, the model still mainly focuses on low frequency spectrograms below 1 kHz. However, the remaining silence in the middle of the speech attracts attention from the model for the speech generated by A09. Therefore, the detection error rate of A09 is lower than that of A10 algorithm without silence at all. 

While for VC algorithms such as A18, the attention of the model trained without silence is distributed over the remaining silence, especially at the onset boundaries of speech. Although VC algorithms retain the proportion of silence duration, the algorithms, except A06 and A19, do not exclude the silence during the conversion. This results in their silence being different from bonafide speech. Because A06 and A19 do not change the non-speech frames \cite{wang2020asvspoof}, most of the silence is the same as bonafide speech. But there are still differences as shown in Figure \ref{fig_sil_content}. Therefore the model trained with the speech after VAD still pays attention to the remaining silence. In addition, due to the similarity of silence, the model mainly detects VC spoof speech through the non-silence low-frequency spectrograms similar to configuration \romannumeral1. So its performance degradation is relatively small compared to detecting TTS spoof speech.

\subsection{Enhancing Robustness by Masking Silence}\label{sub4.3}
As shown in Section \ref{section2}, without additional processing of silence, the proportion of silence duration is an important gap between speech generated by some TTS algorithms and bonafide speech. %And the proportion of silence duration can effectively identify the TTS speech. 
While the spoof speech generated by different waveform generators also differs from the bonafide speech in terms of the content of silence. In the CAM-based visual analysis, even if the silence in the speech is removed through VAD, the model still pays attention to the remaining silence. The second and last heatmap on the right of Figure. \ref{fig_cam} shows that the model assigns attention to the full band of the spectrograms at the remaining silence. To more convincingly demonstrate the focus of the model on remaining silence, an experiment is conducted with the window lengths of the VAD algorithm set to 10ms, 20ms, and 30ms respectively. The larger the window length the lower the time domain resolution of the VAD and the more silence that remains. The results of the experiment are shown in Table \ref{tab:vad_length}.

\begin{table}[htbp]
	\caption{The EER/\% with different VAD window lengths on ASVspoof 2019 LA development (Dev) and evaluation (Eval) datasets. \label{tab:vad_length}}
	\centering
	\begin{tabular}{c c c c c c c}
		\hline
		\multirow{2}*{Window lengths} & \multicolumn{2}{c}{SENet} & \multicolumn{2}{c}{LCNN} & \multicolumn{2}{c}{AASIST} \\
		\cline{2-7}
		& Dev & Eval & Dev & Eval & Dev & Eval \\
		\hline
		VAD 10ms & 3.26 & 20.11 & 13.30 & 26.39 & 1.34 & 20.96 \\
		VAD 20ms & 2.47 & 14.07 & 8.56 & 24.06 & 1.45 & 21.17 \\
		VAD 30ms & 2.87 & 8.73 & 6.87 & 23.12 & 1.25 & 17.54 \\
		\hline
	\end{tabular}
\end{table}

Experimental results show that the EERs of all systems tend to decrease with increasing VAD algorithm window length and increasing remaining silence. This indicates that the anti-spoofing models are concerned about remaining silence. Combined with the analysis of the content of silence in Section \ref{section2}, we hypothesize that silence can be utilized to detect spoof speech. Even though only silence remains, CMs still have the ability to detect spoof speech. %The content of silence proved to be an important basis for CMs and even caused overfitting.% So masking silence can improve the robustness of CM.

\begin{table*}[htbp]
	\caption{Results in terms of EER/\% for algorithms in ASVspoof 2019 LA dev and eval datasets with different masked features. The proportion represents the proportion of silence duration in Section \ref{section2}. \label{tab:mask}}
	\centering
	\setlength{\tabcolsep}{3.0pt}
	\begin{tabular}{ l | c | c c c c c c | c | c c c c c c c c c c c c c | r}
		\hline
		\textbf{Model} & Sil & A01 & A02 & A03 & A04 & A05 & A06 & Dev & A07 & A08 & A09 & A10 & A11 & A12 & A13 & A14 & A15 & A16 & A17 & A18 & A19 & Eval\\
		\hline
		Proportion & - & 1.96 & 46.6 & 12.9 & 6.91 & 50.4 & 47.9 & 31.68 & 0.20 & 5.33 & 2.60 & 0.39 & 0.42 & 2.53 & 0.20 & 1.33 & 2.12 & 7.89 & 47.3 & 46.3 & 46.1 & 18.2 \\
		\hline
		\multirow{3}{*}{SENet} & 1 & 1.18 & 1.05 & 1.05 & 2.35 & 1.37 & 6.05 & 2.98 & 0.51 & 3.91 & 0.41 & 0.65 & 0.55 & 0.55 & 0.86 & 0.63 & 1.28 & 1.75 & 19.8 & 6.22 & 12.8 & 6.24\\
		& 2 & 2.28 & 2.71 & 2.28 & 2.67 & 3.18 & 3.61 & \textbf{2.83} & 0.59 & 6.15 & 0.19 & 0.81 & 0.96 & 0.90 & 0.57 & 0.55 & 0.68 & 0.90 & 5.66 & 5.15 & 9.60 & \textbf{3.44} \\
		& 3 & 0.24 & 0.51 & 0.24 & 0.48 & 0.54 & 1.02 & 0.55 & 0.19 & 1.77 & 0.06 & 0.37 & 0.53 & 0.08 & 0.16 & 0.08 & 0.27 & 0.35 & 2.50 & 1.24 & 1.48 & 1.14\\
		\hline
		\multirow{3}{*}{LCNN} & 1 & 5.89 & 0.11 & 2.71 & 17.5 & 1.15 & 26.8 & 12.8 & 4.36 & 17.7 & 2.99 & 4.50 & 3.66 & 6.86 & 3.62 & 6.24 & 7.91 & 12.3 & 40.5 & 37.5 & 42.3 & 18.5 \\
		& 2 & 0.48 & 1.42 & 0.19 & 0.99 & 0.91 & 1.34 & 1.02 & 0.88 & \textbf{0.99} $\downarrow$ & \textbf{0.02} $\downarrow$ & 0.95 & 0.95 & 0.83 & 0.75 & 0.77 & 0.63 & 1.24 & \textbf{6.33} $\downarrow$ & 20.0 & 6.68 & \textbf{4.66} $\downarrow$ \\
		& 3 & 0.40 & 1.25 & 0.08 & 1.13 & 0.35 & 1.41 & 0.94 & 0.51 & 3.15 & 0.08 & 0.55 & 0.53 & 0.47 & 0.47 & 0.42 & 0.38 & 0.71 & 21.9 & 17.6 & 6.63 & 6.76 \\
		\hline
		\multirow{3}{*}{AASIST} & 1 & 0.11 & 0.00 & 0.03 & 0.24 & 0.00 & 0.24 & 0.15 & 1.97 & 12.8 & 0.45 & 4.23 & 4.25 & 1.77 & 0.79 & 4.70 & 11.2 & 5.45 & 45.3 & 7.20 & 14.4 & 11.2 \\
		& 2 & 0.24 & 0.16 & 0.13 & 0.82 & 0.30 & 1.64 & \textbf{0.71} & 1.18 & 1.16 & 0.02 & 1.18 & 0.31 & 1.22 & 0.11 & 0.53 & 1.20 & 1.63 & 3.17 & 7.43 & 1.02 & \textbf{2.50} \\
		& 3 & 0.24 & 0.16 & 0.11 & 0.62 & 0.40 & 0.94 & 0.47 & 0.91 & 0.38 & 0.04 & 1.24 & 0.45 & 0.87 & 0.14 & 0.20 & 0.67 & 0.95 & 1.69 & 4.56 & 0.67 & 1.59 \\
		\hline
	\end{tabular}
	
\end{table*}

There are three configurations as follows: 
\begin{enumerate}[label={\arabic*.}]
	%\item The models training and prediction are done using the speech cut out of the silence by VAD.
	\item After the silence frames are determined by VAD, the speech frames can be masked (Speech-mask), and only the duration, position and content of silence can be retained. In this way, it is possible to detect whether the spoof speech can be detected only by silence.
	\item Conversely, the silence frames are masked by zero (Sil-mask). Sil-mask preserves the duration and position of silence and removes the content of silence to explore the impact of the content of silence on CMs.
	\item The models training and prediction are done using the unprocessed speech.
\end{enumerate}

The experimental results are shown in Table \ref{tab:mask}. TTS-generated spoof speech can be effectively detected using nothing but silence in configuration 1, even when the speech is masked. The content of silence also contributes to the detection of VC speech compared to using only the proportion of silence duration. The comparison of the experimental results between configuration 2 and 3 provides a more direct understanding of the role played by the content of silence in CMs. After masking the silence, the EER of the SENet model increases by more than 4x and 2x on the development and evaluation partitions, respectively. While the EER of AASIST is increased by 51\% and 57\%, respectively. The lack of the content of silence mainly leads to a decrease in the detection of spoof speech generated by VC algorithms. As analyzed above, most VC algorithms change the content of silence. %The impact on the detection of the TTS algorithms is slighter.

The content of silence is proved to be an important basis for CMs. The SENet and AASIST perform worse when silence is masked, especially when detecting spoof speech generated by VC algorithms. However, the EER of the LCNN model on the evaluation partition is relatively decreased by 31\%. %First of all, the LCNN is the least robust model.% in terms of robustness. 
The Sil-mask can improve the performance of the LCNN on evaluating partitions, especially for unknown algorithms such as A08 and A17. This is because LCNN models with lower robustness have trouble concentrating on the details of features. On the one hand, the model can still detect spoof speech produced by TTS algorithms after masking silence depending on the proportion of silence duration. To identify spoof speech generated by VC algorithms, on the other hand, the model can pay more attention to the non-silence parts of speech. Therefore, we argue that the model may overfit the silence in the training set because the silence parts are constrained by the generation methods inside the dataset. And training models using speech with silence masked can improve the robustness of models with a poor performance against unknown algorithms.

\begin{table}[ht]
	\caption{EER/\% for ASVspoof 2021 DF (DF) and ASVspoof 2021 LA (LA) progress and evaluation datasets based on silence masking. \label{tab:mask_sil}}
	\centering
	\begin{tabular}{c c c c c c}
		\hline
		\multirow{2}*{Model} & \multirow{2}*{Silence mask or not} & \multicolumn{2}{c}{DF} & \multicolumn{2}{c}{LA}\\
		\cline{3-6}
		& & Prog & Eval & Prog & Eval \\
		\hline
		\multirow{2}{*}{SENet} & Sil-mask & 6.54 & \textbf{22.14} &7.52 & \textbf{7.93} \\
		& Unprocessed & 7.58 & 22.95 & 9.01 & 8.71 \\
		\hline
		\multirow{2}{*}{LCNN} & Sil-mask & 5.44 & \textbf{20.41} & 6.97 & 7.22 \\
		& Unprocessed & 6.71 & 22.46 & 6.62 & 6.89 \\
		\hline
		\multirow{2}{*}{AASIST} & Sil-mask & 3.61 & \textbf{16.06} & 9.61 & 7.89 \\
		& Unprocessed & 3.71 & 16.92 & 8.65 & 7.52 \\
		\hline
	\end{tabular}
\end{table}

The results on ASVspoof 2021 DF and LA evaluation datasets are shown in Table \ref{tab:mask_sil}. All three CMs generalize poorly to ASVspoof 2021 DF dataset due to the large differences between the training and evaluation datasets. Thus the performance of the three CMs is improved by silence masking. The EERs of the three systems on ASVspoof 2021 DF evaluation set are relatively reduced by 3.5\%, 9.1\%, and 5.1\%. However, the distribution of ASVspoof 2021 LA dataset is relatively similar to the training set. So the LCNN and AASIST show a slight decrease in performance, with relative increases of 4.8\% and 4.9\% in EER, respectively. But the EER of the SENet model is still decreased by 9.0\% relatively after masking silence.

\begin{table}[ht]
	\caption{EER/\% of silence masking on ASVspoof 2015 and ADD 2023. \label{tab:mask_sil_15add}}
	\centering
	\subfloat[EER/\% of experiments on ASVspoof 2015.]{
		\begin{tabular}{c c c c }
			\hline
			Model & Silence mask or not & 15 Dev & 15 Eval\\
			\hline
			\multirow{2}{*}{SENet} & Sil-mask & 12.12 & \textbf{13.15}\\
			 & Unprocessed & 35.49 & 45.09 \\
			\hline
			\multirow{2}{*}{LCNN} & Sil-mask & 18.88 & \textbf{14.42} \\
			& Unprocessed & 21.04 & 15.13 \\
			\hline
			\multirow{2}{*}{AASIST} & Sil-mask & 2.40 & 2.02 \\
			& Unprocessed & 3.60 & 1.99 \\
			\hline
	\end{tabular}}
	\vfil
	\subfloat[EER/\% on ADD 2023 Track 1.2.]{
		\begin{tabular}{c c c c}
			\hline
			Model & Silence mask or not & Round 1 & Round 2\\
			\hline
			\multirow{2}{*}{LCNN} & Sil-mask & \textbf{20.23} & \textbf{17.69} \\
			 & Unprocessed & 27.31 & 21.58 \\
			\hline
	\end{tabular}}
\end{table}

To further demonstrate the effectiveness of masking silence, we conduct cross-dataset experiments on ASVspoof 2015 \cite{wu2015asvspoof}. The models are trained using ASVspoof 2019 LA training set and evaluated on ASVspoof 2015 development (15 Dev) and evaluation (15 Eval) datasets. Table \ref{tab:mask_sil_15add} (a) demonstrates that masking silence effectively improves the cross-dataset performance of SENet and LCNN. For AASIST, the performance is also improved on 15 Dev, while there is almost no change on 15 Eval. The method is also applied in the STFT-LCNN system submitted to ADD 2023 Challenge \cite{add23} Track 1.2. Track 1.2 of ADD 2023 has two rounds of evaluation and required participants to detect fake speech, specifically the speech generated by other participants. Table \ref{tab:mask_sil_15add} (b) indicates that masking silence is still effective on datasets other than ASVspoof and improves the performance by 26\% and 18\% in the first and second rounds of evaluation, respectively.

\section{Attacking countermeasures with silence}\label{section5}
The proportion of silence duration and content of silence are important bases for NN-based CMs to detect spoof speech as shown in Section \ref{section2} and \ref{section4}. On the one hand, although the general speech systems remove the silence through VAD, the performance of most current CMs detecting speech after removing silence will deteriorate \cite{liu2022asvspoof}. So silence attacks are currently easy to implement and difficult to defend against. On the other hand, silence attacks can effectively improve the success rate of spoof speech against detection systems, such as the top two of ADD 2022 track 3.1 \cite{9746164, 10.1145/3552466.3556532}. As spoof speech generation and detection continue to evolve, generation and countermeasures of silence are issues that need to be addressed. In this section, three types of attacks against speech anti-spoofing CMs are proposed and their attack performance is compared. 
%It can be demonstrated that CMs have difficulty detecting spoof speech concatenated with silence at the beginning and end. It can also be found that CMs with two-dimensional input features have different strategies for detecting spoof speech generated by TTS and VC algorithms. 
Only spoof speech is concatenated with silence for two reasons. First of all, the attacker should only deal with spoof speech, and it is debatable whether the post-processed bonafide speech is still bonafide speech. Secondly, the bonafide speech is not changed, so only the changes in spoof speech cause changes in EERs. This is convenient for comparison and analysis of the experimental results.

\subsection{Attacking Countermeasures with Bonafide Silence}
Bonafide silence segments are obtained from the bonafide speech in ASVspoof 2019 LA training set. The VAD is applied to the bonafide speech, and then the first and last segments of silence are saved separately. For a spoof speech, two segments of silence at the beginning and end are randomly selected and concatenated at the beginning and end of the spoof speech. The concatenated spoof speech is saved for evaluation in order to ensure that the experimental results of different CMs are comparable. The spoof speech in ASVspoof 2019 LA development and evaluation partitions, as well as ASVspoof 2021 LA and DF datasets is concatenated with silence segments. Three systems trained with unprocessed speech are attacked.

\begin{table}[ht]
	\caption{EER/\% of attacking CMs with bonafide silence.} \label{tab:attack_bona}
	\centering
	\setlength{\tabcolsep}{5pt}
	\begin{tabular}{c c c c c c c c}
		\hline
		\multirow{2}*{Model} & \multirow{2}*{Setting}& \multicolumn{2}{c}{2019 LA}  & \multicolumn{2}{c}{2021 DF} & \multicolumn{2}{c}{2021 LA}\\
		\cline{3-8}
		 & & Dev & Eval & Prog & Eval & Prog & Eval \\
		\hline
		\multirow{3}{*}{SENet} & Original &  0.55 & 1.14 & 7.58 & 22.95 & 9.01 & 8.71\\
		 & Attack & 13.96 & 23.05 & 44.97 & 72.67 &  47.97 & 58.93 \\
		 & Change & 13.41 & \textbf{21.91} & 37.39 & \textbf{49.72} & 38.96 & \textbf{50.22} \\
		\hline
		\multirow{3}{*}{LCNN} & Original & 0.94 & 6.76 & 6.71 & 22.46 & 6.62 & 6.89 \\
		& Attack & 9.89 & 25.37 & 36.30 & 52.73 & 37.23 & 36.72\\
		& Change & 8.95 & 18.61 & 29.59 & 30.27 & 30.61 & 29.83\\
		\hline
		\multirow{3}{*}{AASIST} & Original & 0.47 & 1.59 & 3.71 & 16.92 & 8.65 & 7.52 \\
		& Attack & 4.63 & 15.88 & 22.76 & 38.41 & 26.05 & 26.55 \\
		& Change & 4.16 & \textbf{14.29} & 19.05 & \textbf{21.49} & 17.40 & \textbf{19.03}\\
		\hline
	\end{tabular}
\end{table}

Table \ref{tab:attack_bona} shows the results obtained by attacking the anti-spoofing systems with bonafide silence. 'Original' represents the unattacked results in Table \ref{tab:mask} and Table \ref{tab:mask_sil}, 'Attack' denotes the results of attacked systems, and 'Change' refers to the difference between EER after and before the attack. The EERs with the largest and smallest changes in each of the three evaluation datasets are bolded. Since only the spoof speech is manipulated, the increase of EER is mainly due to the false acceptance of spoof speech. The EERs are greatly increased by the attack based on concatenating bonafide silence across all CMs and datasets. Among all datasets, the 19 LA datasets are relatively less affected. Although the EER of the 21 LA dataset is about 10\% lower than the 21 DF dataset, the increase in EER caused by the bonafide silence attack is similar. %It seems possible that silence becomes more important as a basis for CMs since the evaluation datasets of ASVspoof 2021 do not match the training datasets. %After concatenating bonafide silence at start and end of the spoof speech, a large number of spoof speech is wrongly judged. 
Comparing the impact of bonafide silence attack on different CMs, the AASIST suffers the smallest impact with a maximum increase in EER of 21\%. The SENet is the most vulnerable to bonafide silence attack, with up to 50\% increase in EER. Combined with the results in Table \ref{tab:VAD}, the AASIST is the most robust to the impact of silence. On the one hand, the performance of NN and the ability to fit data improves with the increasing number of parameters \cite{golubeva2021are}. And the robustness decreases with the increase of the width and depth of the over-parameterized network \cite{NEURIPS2022_ea5a63f7}. On the other hand, according to the experimental results, the larger the parameters of the models \cite{9747766}, the higher the EER after being attacked by bonafide silence. Therefore, one possible reason is that the AASIST model has the smallest number of parameters. In addition, as the training method of AASIST selects segments from random points instead of from the starting point, the frequency of silence occurrences during training is reduced and the robustness is improved. The amount of parameters of the LCNN model is between the two, so its robustness to bonafide silence attack is also moderate. In contrast, the SENet model has a large number of parameters. And its input features are STFT spectrograms, which are the largest among the three models. %Furthermore, as can be seen from the CAMs in Figure \ref{fig_cam}, the SENet mainly detects spoof speech generated by TTS algorithms through silence. %From the results in Table \ref{tab:mask}, the other two models pay more attention to non-silence than SENet. 
So the SENet trained with unprocessed speech has the best performance in 19 LA datasets, but its robustness against silence is poor.

\subsection{Attacking Countermeasures with Spoof Silence}
The implementation of the attack with spoof silence is similar to the bonafide silence attack, except that the silence segments are from spoof speech. According to the distribution of silence duration in spoof speech, the silence segments mainly come from VC-generated spoof speech.

\begin{table}[ht]
	\caption{EER/\% of attacking CMs with spoof silence.}
	\label{tab:attack_spoof}
	\centering
	\setlength{\tabcolsep}{5pt}
	\begin{tabular}{c c c c c c c c}
		\hline
		\multirow{2}*{Model} & \multirow{2}*{Setting}& \multicolumn{2}{c}{2019 LA}  & \multicolumn{2}{c}{2021 DF} & \multicolumn{2}{c}{2021 LA}\\
		\cline{3-8}
		& & Dev & Eval & Prog & Eval & Prog & Eval \\
		\hline
		\multirow{3}{*}{SENet} & Original &  0.55 & 1.14 & 7.58 & 22.95 & 9.01 & 8.71\\
		& Attack & 1.88 & 2.65 & 10.56 & 38.29 & 16.71 & 16.37 \\
		& Change & 1.33 & 1.51 & 2.98 & 15.34 & 7.70 & 7.66 \\
		\hline
		\multirow{3}{*}{LCNN} & Original & 0.94 & 6.76 & 6.71 & 22.46 & 6.62 & 6.89 \\
		& Attack & 1.69 & 5.14 & 11.45 & 22.99 & 7.04 & 7.07\\
		& Change & 0.75 & \textbf{-1.62} & 4.74 & \textbf{0.53} & 0.42 & \textbf{0.18}\\
		\hline
		\multirow{3}{*}{AASIST} & Original & 0.47 & 1.59 & 3.71 & 16.92 & 8.65 & 7.52 \\
		& Attack & 3.18 & 10.13 & 16.16 & 32.65 & 19.93 & 18.89\\
		& Change & 2.71 & \textbf{8.54} & 12.45 & \textbf{15.73} & 11.28 & \textbf{11.37}\\
		\hline
	\end{tabular}
\end{table}

Table \ref{tab:attack_spoof} shows the results obtained by attacking the CMs with spoof silence. The EERs with the largest and smallest changes in each of the three evaluation datasets are bolded. In general, the impact of the spoof silence attack on different datasets is similar to that of the bonafide silence attack, but the increase of EER is lower. % in different datasets and CMs. 
The 21 DF is still the most impacted, with the greatest increase in EER by 15.73\%. However, the impact of different attacks on different CMs varies significantly. The AASIST becomes the most vulnerable CM to the attack with spoof silence, while the LCNN is hardly affected. On the 19 LA evaluation partition, the EER of the LCNN system is even reduced by 1.62\%. 

\begin{table*}[th]
	\caption{EER/\% results for algorithms in ASVspoof 2019 LA datasets using unprocessed speech (None), concatenating bonafide silence attacking (Bona), concatenating spoof silence attacking (Spoof) and concatenating white noise attacking (White) \label{tab:attack_ana}.}
	\centering
	\setlength{\tabcolsep}{3.0pt}
	\begin{tabular}{ l | c | c c c c c c | c | c c c c c c c c c c c c c | r}
		\hline
		\textbf{Model} & Attack & A01 & A02 & A03 & A04 & A05 & A06 & Dev & A07 & A08 & A09 & A10 & A11 & A12 & A13 & A14 & A15 & A16 & A17 & A18 & A19 & Eval\\
		\hline
		\multirow{4}{*}{SENet} & None & 0.24 & 0.51 & 0.24 & 0.48 & 0.54 & 1.02 & 0.55 & 0.19 & 1.77 & 0.06 & 0.37 & 0.53 & 0.08 & 0.16 & 0.08 & 0.27 & 0.35 & 2.50 & 1.24 & 1.48 & 1.14\\
		& Bona & 23.7 & 4.39 & 10.4 & 23.1 & 3.06 & 3.45 & 14.0 & 35.5 & 30.3 & 9.81 & 37.9 & 39.2 & 18.3 & 13.4 & 18.6 & 31.8 & 19.1 & 5.80 & 3.75 & 3.40 & 23.1 \\
		& Spoof & 3.06 & 0.54 & 1.34 & 3.14 & \textbf{0.51 $\downarrow$} & \textbf{0.02 $\downarrow$} & 1.88 & 3.97 & 3.40 & 1.02 & 3.97 & 4.72 & 2.16 & 1.40 & 2.26 & 3.82 & 2.34 & \textbf{1.45 $\downarrow$} & \textbf{0.81 $\downarrow$} & \textbf{0.90 $\downarrow$} & 2.65 \\
		& White & 2.94 & 0.32 & 0.83 & 4.31 & \textbf{0.43 $\downarrow$} & \textbf{0.70 $\downarrow$} & 2.67 & 3.54 & 2.22 & 0.90 & 4.24 & 4.78 & 2.62 & 1.39 & 1.83 & 3.40 & 2.56 & \textbf{1.20 $\downarrow$} & \textbf{0.61 $\downarrow$} & \textbf{0.71 $\downarrow$} & 2.90 \\
		\hline
		\multirow{4}{*}{LCNN} & None & 0.40 & 1.25 & 0.08 & 1.13 & 0.35 & 1.41 & 0.94 & 0.51 & 3.15 & 0.08 & 0.55 & 0.53 & 0.47 & 0.47 & 0.42 & 0.38 & 0.71 & 21.9 & 17.6 & 6.63 & 6.76 \\
		& Bona & 16.0 & 6.08 & 3.72 & 22.7 & 3.89 & 2.51 & 9.89 & 41.7 & 15.1 & 2.49 & 41.8 & 35.9 & 12.1 & 30.2 & 27.4 & 21.2 & 22.0 & 27.2 & 24.5 & 9.40 & 25.4 \\
		& Spoof & 2.62 & 1.49 & 0.94 & 3.45 & \textbf{0.83 $\downarrow$} & \textbf{0.94 $\downarrow$} & 1.69 & 5.31 & 3.06 & 0.45 & 5.70 & 5.06 & 1.87 & 4.13 & 4.25 & 3.68 & 3.24 & \textbf{10.4 $\downarrow$} & \textbf{9.07 $\downarrow$} & \textbf{3.72 $\downarrow$} & 5.14\\
		& White & 3.45 & 2.57 & 1.92 & 10.6 & 1.57 & 1.85 & 4.23 & 13.4 & 5.06 & 1.24 & 12.8 & 11.3 & 6.25 & 8.31 & 7.29 & 6.01 & 9.26 & \textbf{14.7 $\downarrow$} & \textbf{11.7 $\downarrow$} & \textbf{5.44 $\downarrow$ }& 9.30 \\
		\hline
		\multirow{4}{*}{AASIST} & None & 0.24 & 0.16 & 0.11 & 0.62 & 0.40 & 0.94 & 0.47 & 0.91 & 0.38 & 0.04 & 1.24 & 0.45 & 0.87 & 0.14 & 0.20 & 0.67 & 0.95 & 1.69 & 4.56 & 0.67 & 1.59 \\
		& Bona & 2.47 & 0.48 & 1.05 & 13.4 & 1.05 & 1.45 & 4.63 & 27.5 & 1.91 & 0.55 & 36.4 & 10.9 & 33.5 & 7.00 & 6.51 & 21.39 & 13.6 & 2.44 & 5.76 & 2.26 & 15.9 \\
		& Spoof & 1.61 & 0.27 & 0.59 & 8.75 & 0.83 & 1.29 & 3.18 & 17.2 & 0.96 & 0.33 & 23.3 & 5.97 & 21.8 & 4.91 & 3.46 & 13.2 & 9.30 & 2.32 & 5.19 & 1.87 & 10.1 \\
		& White & 1.37 & 1.02 & 0.48 & 10.8 & 1.80 & 2.00 & 3.92 & 20.0 & 0.95 & 0.20 & 26.7 & 6.73 & 28.2 & 4.58 & 3.54 & 17.5 & 11.4 & 2.32 & 5.23 & 2.36 & 12.6 \\
		\hline
	\end{tabular}
	
\end{table*}

In order to explore the reasons for the changes in EER, the impacts of the two attack methods on different spoofing algorithms are shown in Table \ref{tab:attack_ana}. The EERs of all CMs and spoofing algorithms increase when bonafide silence is concatenated. This suggests that the CMs pay attention to the silence part of speech. The impact of concatenating spoof silence is more complex. The impact on the AASIST-based CM is similar to that of concatenating bonafide silence. However, concatenating spoof silence increases the EERs of TTS algorithms for SENet and LCNN while decreasing the EERs of VC algorithms. CNN-based LCNN and SENet have two-dimensional input features with stronger discrimination ability. Consequently, the detection of TTS algorithms and VC algorithms is affected differently by spoof silence attacks. For the spoof speech generated by TTS algorithms without silence, any type of silence attack can significantly improve the detection EER of CMs. As for the spoof speech generated by VC algorithms, Section \ref{section2} and Subsection \ref{sub4.3} indicate that the content of silence is an effective basis for discrimination. The experimental results here demonstrate from another perspective that the CMs with two-dimensional inputs can identify the artifacts in the silence part of spoof speech generated by VC algorithms. %After being attacked by spoof silence, the EERs of SENet and LCNN are reduced in detecting VC algorithm. 
Since the spoof silence mainly comes from the known spoof speech generated by VC algorithms, the artifacts of the spoof silence are enhanced after concatenating the spoof silence at the beginning and end of the VC spoof speech. And the LCNN and SENet can correct the judgment results of a part of spoof speech generated by VC algorithms, % that is originally misjudged in the test sets, 
especially A17 which does not change silence. %The AASIST is robust to the content of silence relatively. Although it has a stronger robustness against the bonafide silence attack, it is more vulnerable to spoof silence attacks.

\subsection{Attacking Countermeasures with White Noise}
The two attacks mentioned above % that use bonafide or spoof silence 
require silence data and are probably slightly difficult to implement. Therefore, an attack with white noise, which is easier to implement is proposed. Since the average signal-to-noise ratio (SNR) of the speech in ASVspoof 2019 training set is about 40 dB, two segments of white noise with an SNR of 40 dB are concatenated at the beginning and end of the spoof speech. Because the average duration of silence in bonafide speech is about $2/5$ of the whole speech, the length of each segment of silence is evenly distributed in $(0, 0.4 * T)$ where $T$ is the duration of the whole speech. %The post-processing is also conducted offline and the post-processed spoof speech is saved, too. 

\begin{table}[ht]
	\caption{EER/\% of attacking CMs with white noise.}
	\label{tab:attack_white}
	\centering
	\setlength{\tabcolsep}{5pt}
	\begin{tabular}{c c c c c c c c}
		\hline
		\multirow{2}*{Model} & \multirow{2}*{Setting}& \multicolumn{2}{c}{2019 LA}  & \multicolumn{2}{c}{2021 DF} & \multicolumn{2}{c}{2021 LA}\\
		\cline{3-8}
		& & Dev & Eval & Prog & Eval & Prog & Eval \\
		\hline
		\multirow{3}{*}{SENet} & Original &  0.55 & 1.14 & 7.58 & 22.95 & 9.01 & 8.71\\
		& Attack & 2.67 & 2.90 & 12.17 & 45.60 &  21.71 & 20.60 \\
		& Change & 2.12 & 1.76 & 4.59 & 22.65 & 12.70 & 11.89 \\
		\hline
		\multirow{3}{*}{LCNN} & Original & 0.94 & 6.76 & 6.71 & 22.46 & 6.62 & 6.89 \\
		& Attack & 4.23 & 9.30 & 18.12 & 45.52 & 13.13 & 13.01\\
		& Change & 3.29 & 2.54 & 11.41 & \textbf{23.06} & 6.51 & 6.12\\
		\hline
		\multirow{3}{*}{AASIST} & Original & 0.47 & 1.59 & 3.71 & 16.92 & 8.65 & 7.52 \\
		& Attack & 3.92 & 12.59 & 18.03 & 36.47 & 22.18 & 21.32 \\
		& Change & 3.45 & \textbf{11.00} & 14.32 & 19.55 & 13.53 & \textbf{13.80}\\
		\hline
	\end{tabular}
\end{table}

The experimental results are shown in Table \ref{tab:attack_white}, the EERs with the largest changes in each of the three evaluation datasets are bolded. Compared with the results in Table \ref{tab:attack_bona} and Table \ref{tab:attack_spoof}, white noise attacks result in an increase in EER that is more than attacks with spoof silence and less than attacks that use bonafide silence. %The impact of attacks on different datasets and CMs is similar to other attacks. 
The 21 DF suffers the greatest impact while the 19 LA suffers the least. According to the results in Table \ref{tab:attack_ana}, although the white noise attack can also increase the EERs of TTS algorithms, the value is lower since it differs from bonafide silence. White noise, however, is analogous to bonafide silence rather than spoof silence processed by spoofing algorithms. Thus for SENet and LCNN, the EER increase of TTS algorithms caused by white noise is greater than spoof silence, while the EER reduction of VC algorithms is lower. The AASIST is the most vulnerable system to white noise attack. The EERs of all algorithms are also increased, but the value is between the increases caused by the bonafide silence attack and spoof silence attack.

In summary, the spoof speech detection systems can be attacked by concatenating silence at the beginning and end of the spoof speech, specifically to increase the confusion of the unidentified TTS spoofing algorithms.

\section{Mitigate the impact of silence}
Recently there is work related to enhancing the robustness of CMs to VAD, such as low-pass filtering and bandwidth extension \cite{wang2022low} and pre-trained model \cite{wang2023spoofed}. However, these methods are complex and there is still a lack of methods against the silence attacks. It is demonstrated that the spectrogram in low-frequency part below 4 kHz has better performance and robustness than the spectrogram in full-frequency in our previous work \cite{zhang21da_interspeech}. Moreover, the CAMs in Figure \ref{fig_cam} show that the SENet-based CM mainly concentrates on spectrograms below 1 kHz. Similarly, the champion of ASVspoof 2021 LA and DF tasks used a low-pass filter to improve robustness, but with a cutoff frequency of 3.4 kHz~\cite{tomilov21_asvspoof}. The speech below 1 kHz mainly describes information such as fundamental frequency (\textit{F0}), lower formant, and pitch. The low-frequency information can improve the performance of speech recognition \cite{chang2006unintelligible, luo2006contribution}, and enhance the \textit{F0} and amplitude envelope representations of vocoders \cite{qin2006effects, brown2009low}. Pitch and \textit{F0} are also important indicators to examine the performance of TTS and VC algorithms \cite{ren2020fastspeech, qian2020f0}. Therefore, the performance and robustness of low-frequency information below 1 kHz in speech anti-spoofing are explored in this section. The experimental results show that using only low-frequency information can effectively improve the robustness of CMs against VAD and silence attacks.%, regardless of whether it is VAD or attack with concatenating silence.

\subsection{Mitigate the Impact of VAD}
\begin{figure}[ht]
	\centering
	\includegraphics[width=1.0\linewidth]{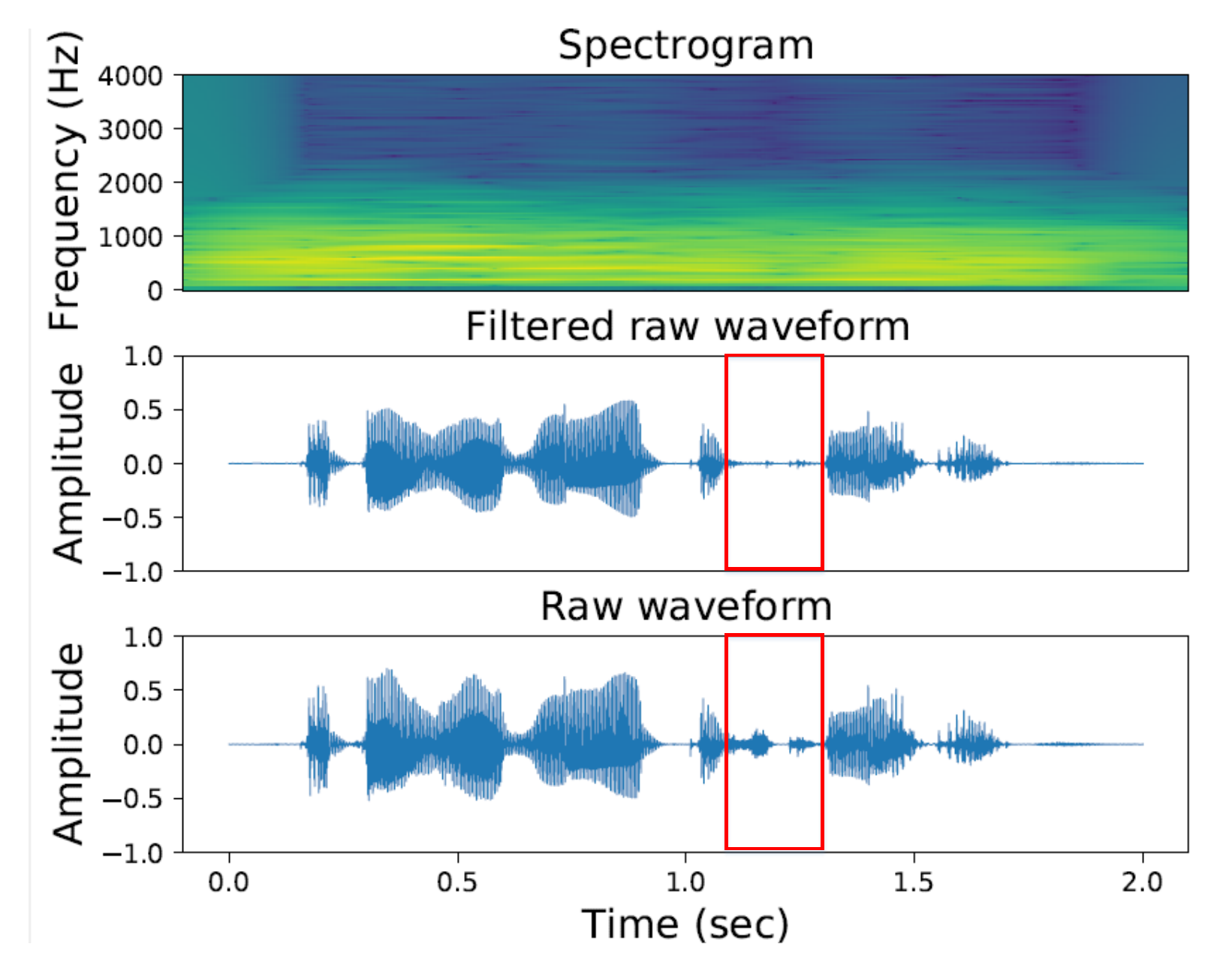}
	\caption{The waveforms of speech before and after being filtered, as well as the spectrogram feature of filtered speech.}
	\label{fig_filter}
\end{figure}

Many speech signal processing systems remove the silence by VAD before feeding the speech. But current spoof speech detection systems have poor performance after VAD. So mitigating the impact of VAD is important for the practical application of CMs. According to the results in Section \ref{section4}, the model trained with speech without silence can get better performance in the face of the test speech after VAD. So speech filtered by a low-pass filter with a cutoff frequency of 1 kHz similar to \cite{tomilov21_asvspoof} is used to train the CMs. The low-pass filter is obtained through the 8th-order Butterworth filter. The filtered speech waveform and spectrogram are shown in Figure \ref{fig_filter}. For the waveform, the high-frequency details such as those inside the red box are lost after filtering. %, and the signal at longer time scales is retained. 
And for the spectrogram, the high-frequency part is blurred and only the details below the cutoff frequency are retained.

The speech after filtering and VAD is used to retrain the three CMs in Section \ref{section3}. Their performance in detecting spoof speech without silence is evaluated as shown in table \ref{tab:1k_vad}. The smallest EERs for the three evaluation datasets and the most reduced EERs for the three models are bolded separately. The robustness of the CMs to spoof speech without silence, especially the spoof speech generated by unknown algorithms, can be effectively improved by training and testing with the speech below 1 kHz obtained by low-pass filtering. Compared with unfiltered speech, the EERs are reduced by about 2\% to 9\% on 19 LA evaluation partition, as well as 21 LA and 21 DF datasets containing unknown spoofing algorithms and channel interference outside the training set. However, there is an increase in EER for 19 LA development partition containing known spoofing algorithms. Among the three CMs, the AASIST still has the best performance.

\begin{table}[ht]
	\caption{EER\% obtained from speech after low-pass filtering and VAD, and the change of EER compared with the results obtained by speech after VAD in Tables \ref{tab:VAD} and Table \ref{tab:vad_21}.}
	\label{tab:1k_vad}
	\centering
	\setlength{\tabcolsep}{5pt}
	\begin{tabular}{c c c c c c c c}
		\hline
		\multirow{2}*{Model} & \multirow{2}*{Setting}& \multicolumn{2}{c}{2019 LA}  & \multicolumn{2}{c}{2021 DF} & \multicolumn{2}{c}{2021 LA}\\
		\cline{3-8}
		& & Dev & Eval & Prog & Eval & Prog & Eval \\
		\hline
		\multirow{3}{*}{SENet} & VAD & 3.26 & 20.11 & 28.35 & 30.89 & 26.49 & 28.31\\
		& Filtered & 6.04 & \textbf{18.14} & 21.27 & 25.73 &  24.28 & 24.95 \\
		& Change & 2.78 & -1.97 & \textbf{-7.08} & -5.16 & -2.21 & -3.36 \\
		\hline
		\multirow{3}{*}{LCNN} & VAD & 13.30 & 26.39 & 30.24 & 33.23 & 31.75 & 32.96 \\
		& Filtered & 10.20 & 23.34 & 21.15 & 28.88 & 24.46 & 24.25 \\
		& Change & -3.10 & -3.05 & \textbf{-9.09} & -4.35 & -7.29 & -8.71 \\
		\hline
		\multirow{2}{*}{AASIST} & VAD & 1.13 & 21.0 & 25.64 & 27.88 & 26.13 & 27.66 \\
		& Filtered & 2.47 & 18.93 & 18.74 & \textbf{19.23} & 23.57 & \textbf{21.59} \\
		& Change & 1.13 & -2.03 & -6.90 & \textbf{-8.65} & -2.56 & -6.07\\
		\hline
	\end{tabular}
\end{table}

\subsection{Mitigate the Impact of Silence Attacks}
In order to confront different types of silence attacks, a method of speech anti-spoofing fusing the results by silence and speech separately is proposed. The silence frames and speech frames are first distinguished by the VAD algorithm. Then two spoof speech detection models are trained using full-band silence and 1 kHz low-pass filtered speech respectively. Finally, the scores of the two systems are fused in equal proportions to obtain the final decision. The experimental results of the proposed method on different datasets for silence attacks are as follows and the smallest EERs for the three evaluation datasets and the most reduced EERs for the three models are bolded separately:

\begin{table}[htbp]
	\caption{EER\% obtained from the proposed method and the change of EER compared with the results obtained by attacking the system with bonafide silence in Table \ref{tab:attack_bona}, spoof silence in Table \ref{tab:attack_spoof} and white noise in Table \ref{tab:attack_white}.}
	\label{tab:sil1k_attack}
	\centering
	\setlength{\tabcolsep}{4.2pt}
	\subfloat[Mitigate bonafide silence attacks.]{
		\begin{tabular}{c c c c c c c c}
			\hline
			\multirow{2}*{Model} & \multirow{2}*{Setting}& \multicolumn{2}{c}{2019 LA}  & \multicolumn{2}{c}{2021 DF} & \multicolumn{2}{c}{2021 LA}\\
			\cline{3-8}
			& & Dev & Eval & Prog & Eval & Prog & Eval \\
			\hline
			\multirow{3}{*}{SENet} & Attack & 13.96 & 23.05 & 44.97 & 72.67 &  47.97 & 58.93\\
			& Mitigated & 2.62 & 16.04 & 25.03 & \textbf{29.01} & 28.94 & 29.45 \\
			& Change & -11.34 & -7.01 & -19.94 & \textbf{-43.66} & -19.03 & -29.48 \\
			\hline
			\multirow{3}{*}{LCNN} & Attack & 9.89 & 25.37 & 36.30 & 52.73 & 37.23 & 36.72 \\
			& Mitigated & 14.63 & 29.37 & 26.70 & 35.82 & 24.63 & \textbf{25.22}\\
			& Change & 4.74 & 4.00 & -9.60 & \textbf{-16.91} & -12.60 & -11.50\\
			\hline
			\multirow{3}{*}{AASIST} & Attack & 4.63 & 15.88 & 22.76 & 38.41 & 26.05 & 26.55 \\
			& Mitigated & 1.57 & \textbf{15.69} & 18.22 & 36.01 & 30.43 & 32.62 \\
			& Change & -3.06 & -0.19 & \textbf{-4.54} & -2.40 & 4.38 & 6.07\\
			\hline
		\end{tabular}}
	\vfil
	\subfloat[Mitigate spoof silence attacks.]{
		\begin{tabular}{c c c c c c c c}
			\hline
			\multirow{2}*{Model} & \multirow{2}*{Setting}& \multicolumn{2}{c}{2019 LA}  & \multicolumn{2}{c}{2021 DF} & \multicolumn{2}{c}{2021 LA}\\
			\cline{3-8}
			& & Dev & Eval & Prog & Eval & Prog & Eval \\
			\hline
			\multirow{3}{*}{SENet} & Attack & 1.88 & 2.65 & 10.56 & 38.29 & 16.71 & 16.37 \\
			& Mitigated & 0.04 & 1.33 & 17.39 & 22.62 & 20.83 & 22.03 \\
			& Change & -1.84 & -1.32 & 6.83 & \textbf{-15.67} & 4.12 & 5.66 \\
			\hline
			\multirow{3}{*}{LCNN} & Attack & 1.69 & 5.14 & 11.45 & 22.99 & 7.04 & 7.07 \\
			& Mitigated & 3.96 & 8.66 & 12.88 & 19.61 & 8.77 & \textbf{9.21}\\
			& Change & 2.27 & 3.52 & 1.43 & \textbf{-3.38} & 1.73 & 2.14\\
			\hline
			\multirow{3}{*}{AASIST} & Attack & 3.18 & 10.13 & 16.16 & 32.65 & 19.93 & 18.89 \\
			& Mitigated & 0.01 & \textbf{0.99} & 1.42 & \textbf{10.95 }& 15.51 & 20.48 \\
			& Change & -3.17 & -9.14 & -14.74 & \textbf{-21.70} & -4.42 & 1.59\\
			\hline
			\end{tabular}}
	\vfil
	\subfloat[Mitigate white noise attacks.]{
		\begin{tabular}{c c c c c c c c}
			\hline
			\multirow{2}*{Model} & \multirow{2}*{Setting}& \multicolumn{2}{c}{2019 LA}  & \multicolumn{2}{c}{2021 DF} & \multicolumn{2}{c}{2021 LA}\\
			\cline{3-8}
			& & Dev & Eval & Prog & Eval & Prog & Eval \\
			\hline
			\multirow{3}{*}{SENet} & Attack & 2.67 & 2.90 & 12.17 & 45.60 &  21.71 & 20.60\\
			& Mitigated & 0.04 & \textbf{0.53} & 16.44 & \textbf{18.66} & 19.64 & 19.96 \\
			& Change & -2.63 & -2.37 & 4.27 & \textbf{-26.94} & -2.07 & -0.64 \\
			\hline
			\multirow{3}{*}{LCNN} & Attack & 4.23 & 9.30 & 18.12 & 45.52 & 13.13 & 13.01 \\
			& Mitigated & 5.66 & 9.72 & 18.93 & 26.69 & 13.54 & \textbf{14.17} \\
			& Change & 1.43 & 0.42 & 0.81 & \textbf{-18.83} & 0.41 & 1.16\\
			\hline
			\multirow{3}{*}{AASIST} & Attack & 3.92 & 12.59 & 18.03 & 36.47 & 22.18 & 21.32 \\
			& Mitigated & 0.43 & 3.71 & 6.44 & 25.77 & 23.87 & 19.24 \\
			& Change & -3.49 & -8.88 & -11.59 & \textbf{-10.70} & 1.69 & -2.08\\
			\hline
			\end{tabular}}
\end{table}

As shown in Table \ref{tab:sil1k_attack}(a), detecting spoof speech with full-band silence and low-frequency speech respectively can mitigate the impact of the bonafide silence attack. The EERs decrease in almost all cases except for the LCNN system on 19 LA datasets, as well as the AASIST on 21 LA datasets. The robustness of the SENet against bonafide silence attacks has been greatly improved.

In contrast, for the attack with spoof silence and white noise, the proposed method has uneven mitigation effects on different datasets and systems, as shown in Table \ref{tab:sil1k_attack} (b) and (c). The possible causes are that the content of silence is an important judgment basis for CMs, and silence attacks are only performed on spoof speech. As a result, the more discriminative SENet and AASIST effectively defended against these two silence attacks on 19 LA and 21 DF datasets.

Additionally, most systems under silence attack still have an EER of around 20\% on ASVspoof 2021 datasets. This is a big gap compared to SOTA anti-spoofing systems.

\section{Conclusion and Discussion}
In recent years, various NN-based speech anti-spoofing CMs have significantly improved the performance of spoof speech detection on specific datasets such as ASVspoof. However, most of the CMs lack interpretability and are not robust enough to be applied to practical data. In particular, silence has been shown to have a significant impact on the performance of speech anti-spoofing CMs. The EER of CMs increases significantly when silence is removed by VAD. There are still some questions that remain unclear, such as whether CMs are more concerned about silence or speech, and why? Therefore, it is of great significance to investigate the reasons for the impact of silence on CMs and explore methods to mitigate the performance degradation caused by removing and concatenating silence.

In this paper, the reasons why silence impacts speech anti-spoofing systems are analyzed. On the one hand, TTS algorithms generate spoof speech with a low proportion of silence duration when the silence is not processed. The spoof speech generated by TTS algorithms can be distinguished even if the proportion of silence duration is directly used as scores in ASVspoof 2019 LA. On the other hand, it can be observed from the spectrograms that spoof speech and bonafide speech differ in the content of silence. This may be caused by differences in the waveform generator.

In order to demonstrate these findings, experiments are performed on three CMs with various silence configurations. The different impact of silence on the spoofing algorithms is explained in terms of the difference in the theory of the spoofing algorithms. And the models trained with unprocessed speech tend to misclassify bonafide speech without silence. While the models trained with speech after VAD tend to falsely accept spoof speech generated by unknown spoofing algorithms. Based on the attention visualization of the SENet through CAM, it appears that the models focus on the region of the low-frequency part below 1 kHz in the speech spectrogram.

For the speech generated by TTS algorithms, the proportion of silence duration becomes one of the important bases for anti-spoofing. % due to the low proportion of silence duration of speech generated by most TTS algorithms. 
As shown in Table \ref{tab:VAD}, the current CMs have difficulty in the detection of unknown TTS algorithms if silence is removed by VAD. However, if the proportion of silence duration is preserved, either by using it directly as a score in Figure \ref{fig_portion_sil} or by masking the silence or even masking the speech in Table \ref{tab:mask}, the TTS algorithms can be detected. As for the speech generated by VC algorithms, the model focuses on the parts of speech and silence at the seams because silence has different contents. Silence content is an important basis for CMs, as demonstrated by experiments on masking silence and speech. Furthermore, masking silence as a way to improve the robustness of speech anti-spoofing systems is proposed. %Masking silence can reduce EER by 3.5\% to 9.1\% on ASVspoof 2021 DF evaluation dataset.

Then, attacks on the CMs by concatenating silence at the beginning and end of spoof speech are proposed. The ability of CMs to discriminate the content of silence can be demonstrated again by comparing the attack results of different silence. Low-frequency filtering is proposed to mitigate the impact of silence because the CAM heatmaps indicate that the SENet model focuses on the spectrogram of low-frequency part below 1 kHz, together with the implication of low-frequency information. Experimental results demonstrate that low-pass filtering can mitigate the performance degradation caused by VAD. And fusing the scores of low-frequency speech and full-frequency silence can improve the robustness of the CMs against silence attacks.

In conclusion, the impact of silence on speech anti-spoofing and its reasons are visually explored and demonstrated. However, after silence is removed, it is still challenging for current CMs to effectively detect whether a speech is spoofed, especially when the spoof speech is generated by unknown algorithms. Therefore, developing interpretable CMs and detecting spoof speech generated by unknown algorithms from the non-silence part is of great significance to the speech anti-spoofing community.

\bibliographystyle{IEEEtran}
\bibliography{mybib}

\end{document}